\documentclass[11pt]{article}
\usepackage[final]{colm2026_conference}

\usepackage{latexsym}
\usepackage{microtype}
\usepackage{hyperref}
\usepackage{url}
\usepackage{booktabs}
\usepackage{lineno}

\definecolor{darkblue}{rgb}{0, 0, 0.5}
\hypersetup{colorlinks=true, citecolor=darkblue, linkcolor=darkblue, urlcolor=darkblue}

\usepackage{graphicx}

\usepackage{amsmath,amsfonts,bm}

\def\eqref#1{equation~\ref{#1}}

\def\1{\bm{1}}

\DeclareMathAlphabet{\mathsfit}{\encodingdefault}{\sfdefault}{m}{sl}
\SetMathAlphabet{\mathsfit}{bold}{\encodingdefault}{\sfdefault}{bx}{n}

\DeclareMathOperator{\sign}{sign}

\newcommand\fscore{F$_1$}
\newcommand\semstamp{\texttt{SemStamp} }
\newcommand\ksemstamp{\texttt{$k$-SemStamp} }
\newcommand\kgw{\texttt{KGW} }
\newcommand\sweet{\texttt{SWEET} }
\newcommand\ours{\texttt{SeqMark} }

\usepackage{hyperref}
\usepackage{url}
\usepackage{booktabs}
\usepackage{graphicx}
\usepackage{array}
\usepackage{multirow}
\usepackage{subcaption}  
\usepackage{caption}
\usepackage{wrapfig}
\usepackage[x11names]{xcolor}

\usepackage[ruled,vlined]{algorithm2e}
\usepackage{amsmath}

\title{Semantic Differentiation for Tackling Challenges in Watermarking Low-Entropy Constrained Generation Outputs}

\author{Nghia T. Le, Alan Ritter, \& Kartik Goyal \\
School of Interactive Computing\\
Georgia Institute of Technology\\
Atlanta, GA 30308, USA \\
\texttt{\{nle18,alan.ritter,kartikgo\}@gatech.edu} \\
}

\definecolor{crimsonglory}{rgb}{0.75, 0.0, 0.2}
\definecolor{darkgreen}{rgb}{0.0, 0.5, 0.0}
\definecolor{darkpink}{rgb}{0.8, 0.1, 0.4}

\begin{document}

\maketitle

\ifcolmsubmission
\linenumbers
\fi

\begin{abstract}
We demonstrate that while the current approaches for language model watermarking are effective for open-ended generation, they are inadequate at watermarking LM outputs for constrained generation tasks with low-entropy output spaces. Therefore, we devise \emph{SeqMark}, a sequence-level watermarking algorithm with semantic differentiation that balances output quality, watermark detectability, and imperceptibility.
It improves on the shortcomings of token-level watermarking algorithms that cause under-utilization of the sequence-level entropy available for constrained generation tasks. Moreover, we identify and improve upon the problem of \emph{region collapse}, a different failure mode associated with prior sequence-level watermarking algorithms. This occurs because the pseudorandom partitioning of semantic space for watermarking in these approaches causes all high-probability outputs to collapse into either invalid or valid regions, leading to a trade-off in output quality and watermarking effectiveness. Instead, SeqMark differentiates the high-probable output subspace and partitions it into valid and invalid regions, ensuring the even spread of high-quality outputs among all the regions. On various constrained generation tasks like machine translation, abstractive summarization, and code generation, SeqMark substantially improves watermark detection accuracy (up to 28\% increase in \fscore) while maintaining high generation quality. 
\footnote{Our code is available at \href{https://github.com/nle18/seqmark}{ https://github.com/nle18/seqmark}.}
\end{abstract}

\section{Introduction}
As progress in language modeling leads to increasingly human-like automatic text generation, demand for tracing the provenance and life-cycle of digital text on the internet has soared. Questions around plagiarism~\citep{Info2023ChatGPTIH}, copyright infringement~\citep{copyright}, veracity of text~\citep{augenstein2024factuality}, misinformation~\citep{Gravel2023.03.16.23286914}, multi-agentic communication \citep{motwani2024secret}, and many other legal and contractual frameworks require the ability to attribute text to its source. Language model watermarking -- embedding a traceable digital marker in the language model's distribution over language -- has emerged as a potential solution to address this need. The seminal work on watermarking~\cite{pmlr-v202-kirchenbauer23a} modifies the logits produced over the vocabulary at each generation step from a language model to prefer or disprefer certain tokens.
Much of the follow-up work~\citep{fernandez2023three,Zhao2023ProvableRW, takezawa2025necessarysufficientwatermarklarge} on this \emph{token-level watermarking framework} has focused on improved robustness to edits, generation quality, detectability, and other desiderata for watermarking. While effective for open-ended long-form generation, LM watermarking remains a challenging problem in low-entropy setups such as factual question-answering and code generation. While some work~\citep{lee-etal-2024-wrote, lu-etal-2024-entropy} addresses this problem for code generation via selective token-entropy based watermarking, \emph{we posit that in general, all token-level watermarking schemes under-utilize sequence-level entropy for embedding the signal and hence struggle in constrained generation tasks} like machine translation and abstractive summarization. Although these tasks have lower entropy than open-ended generation, they still admit multiple feasible responses with sufficient sequence-level entropy.

An appealing alternative paradigm is \textit{sequence-level watermarking}~\citep{hou-etal-2024-semstamp, hou-etal-2024-k}: instead of randomly partitioning the token space at each generation step, it focuses on partitioning the manifold of sequences via hyperplanes into red/green regions, and then performs rejection sampling to sample from the green regions. 
However, \emph{we identify a different common issue of red/green region collapse} with existing approaches that prevent low-entropy watermarking. 
Specifically, the operationalization of these algorithms via locality-sensitive hashing or $k$-means clustering leads to the assignment of semantically similar sentences to the same red/green partition. Due to semantic diversity, this is not a problem for open-ended generation tasks, but for constrained generation tasks, most of the admissible outputs with high probability tend to be semantically close to one another and mostly collapse to a single region. This imposes a challenging tradeoff between output quality and watermark verifiability for watermarking in constrained generation tasks.

In this paper, we propose a novel sequence-level watermarking approach \emph{SeqMark} that utilizes the limited sequence entropy effectively to watermark constrained generation tasks. This approach first identifies the semantic manifold of high-probable admissible outputs for the task at hand, and then semantically differentiates between the elements of this manifold to propose pseudorandom red/green partitions in a manner such that the potential high-quality responses are evenly spread across the partitions, enabling successful watermarking in low-entropy settings. We empirically support our claims and observations on watermarking previously underexplored constrained generation tasks like machine translation, summarization, and code generation. We demonstrate the issues with existing token-level and sequence-level watermarking algorithms and show that SeqMark is more effective in constrained generation settings, while maintaining high watermarking capabilities in open-ended generation settings. 
\section{Preliminaries}
\label{sec:prelim}
Watermarking
is inherently a problem of incorporating a digital marker into the (ideally noise-tolerant) carrier signal. 
We are primarily interested in settings that incorporate a marker into the autoregressive language model's distribution and thus perform watermarking~(ideally imperceptible) during text generation. 

\paragraph{Token-level Logit Watermarking} \cite{pmlr-v202-kirchenbauer23a} introduced a seminal token-level watermarking algorithm dubbed KGW that embedded the marker in the logits produced over vocabulary for each token during generation. Many following attempts \citep{liu2024semanticinvariantrobustwatermark, fernandez2023three} proposed variations to improve robustness and undetectability of the watermarking scheme. At an abstract level, all of these approaches modify the \emph{token-level distribution} for watermarking during generation by pseudorandomly partition the vocabulary $\mathcal{V}$ into a green list of size $\gamma|\mathcal{V}|$ and red list size $(1-\gamma)|\mathcal{V}|$ for a hyperparameter $\gamma \in (0,1)$, using a hash from the previous token $w_{t-1}$ or LM state as random seed.
The generation procedure is then modified to prefer the green-list partition over the red-list partition, thus embedding the watermark. Detection involves performing hypothesis testing to decide whether the frequency of the green-list tokens in the query text occurs by chance.

\paragraph{Sequence-level Watermarking} To improve the robustness of the watermarking method to paraphrase attacks, a different line of approach dubbed as semantic watermarking \citep{hou-etal-2024-semstamp, hou-etal-2024-k} focuses on using sequence-level distribution induced by the language model instead of token-level distributions to embed the watermark. 
In general, the \emph{sentence-embedding space} is pseudorandomly partitioned into green and red regions. Then a sentence is generated from the language modeling via rejection sampling until the resulting sentence embedding falls in the green region or the budget for resampling is exhausted. For generating the next sentence, another round of rejection sampling is performed following pseudorandom partitioning of the embeddings space using the hash based on the previously generated sentence. Prior work SemStamp~\citep{hou-etal-2024-semstamp} uses locality-sensitive hashing (LSH) to pseudorandomly partition the embedding space: LSH first samples $n$ random vectors from a normal Gaussian distribution to specify $n$ hyperplanes and thus $2^n$ regions. Given a hyperparameter ratio $\gamma$ and the hash of the previous sentence, the regions are partitioned into $\gamma2^n$ green regions and $(1-\gamma)2^n$ red regions. 
A sentence with embedding $\mathbf{v} \in \mathbb{R}^d$ would receive an $n$-bit binary LSH signature $c$, where each bit specifies the location of $v$ with respect to each hyperplane -- thus $c$ identifies the region assignment for $v$. 
In contrast, $k$-SemStamp \citep{hou-etal-2024-k} uses k-means clustering for partitioning instead of LSH: it first estimates $K$ centroids for a general semantic manifold via pretraining, and then during the generation step it randomly assigns each of these centroids either a red/green tag. During sampling, each generated sentence is assigned red/green tag according to its closest centroid. Watermark detection hinges on the frequency of sentences in the green regions. 

\begin{figure*}[t]
    \centering
    \begin{subfigure}{0.42\textwidth}
        \centering
        \includegraphics[width=\linewidth]{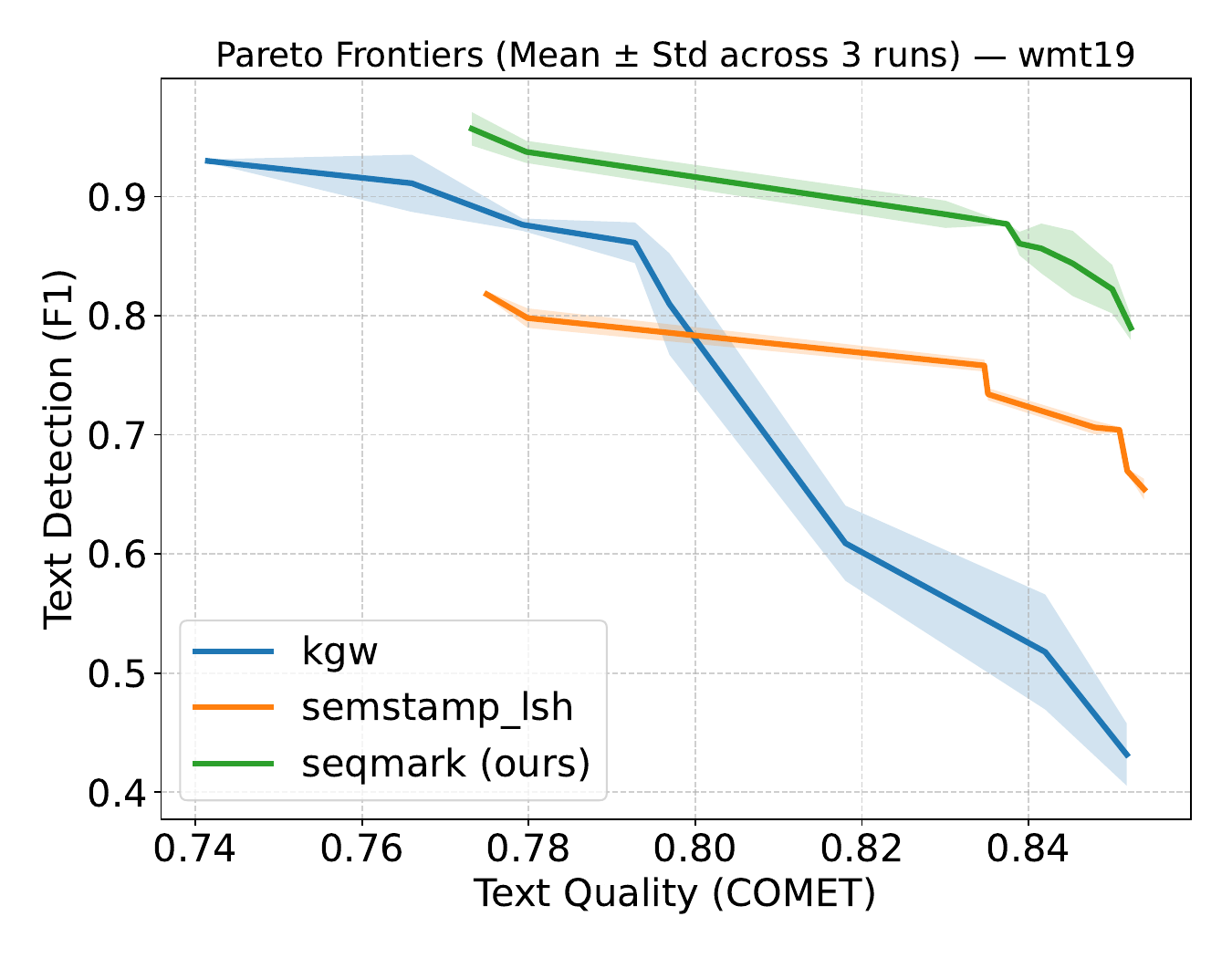} 
    \end{subfigure}
    \begin{subfigure}{0.42\textwidth}
        \centering
        \includegraphics[width=\linewidth]{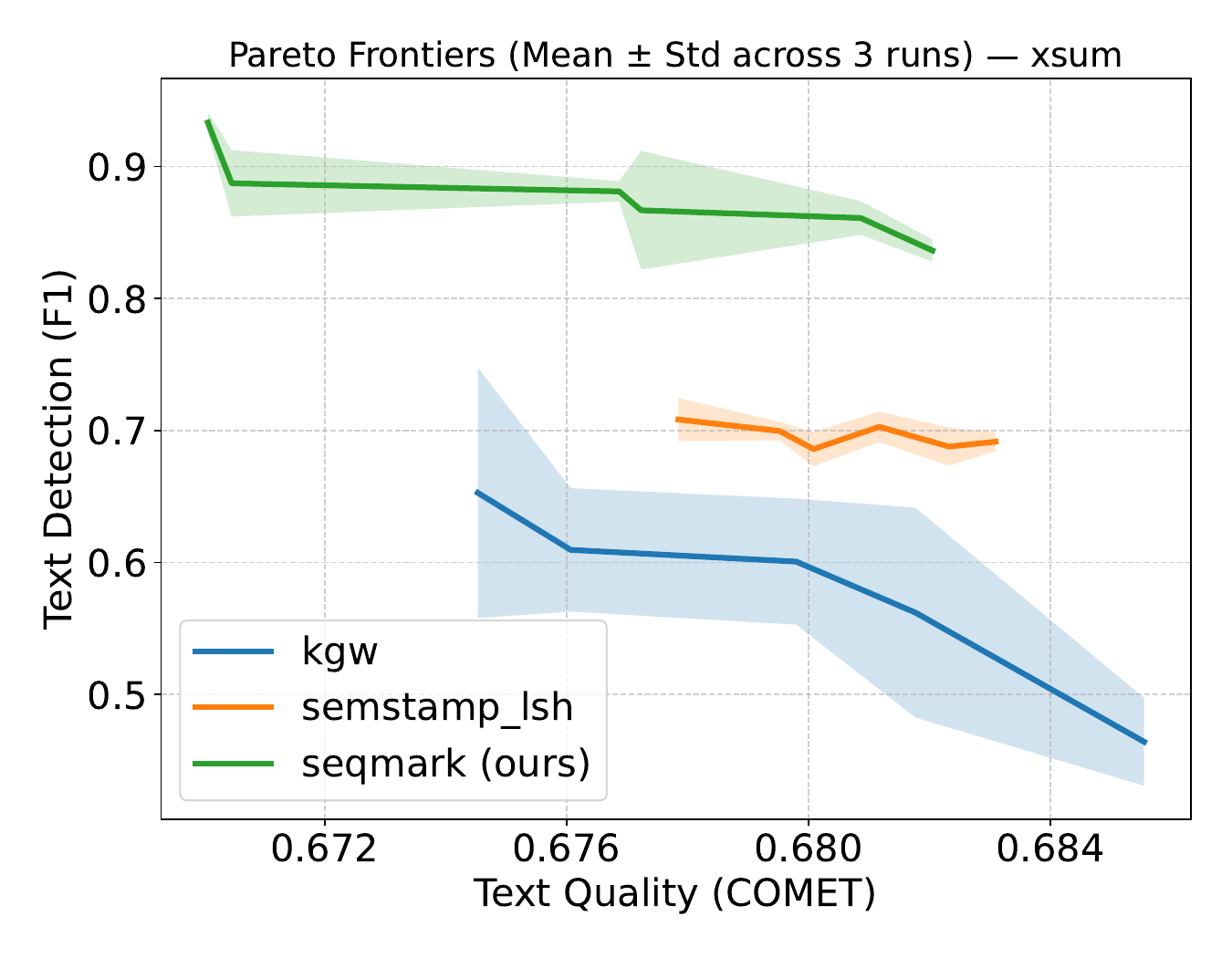}
    \end{subfigure}

    \caption{Pareto frontiers of token-level watermarking KGW (\textcolor{blue}{blue}), and two sequence-level watermarking approaches SemStamp (\textcolor{orange}{orange}) and SeqMark~(ours, \textcolor{darkgreen}{green}) for machine translation (\textbf{Left}) and summarization (\textbf{Right}). 
    }
    \vspace{-5mm}
    \label{fig:pareto}
\end{figure*}

\section{Entropy and Watermarking}

Imperceptibility of the watermark is very important for textual applications in which the flow, tone, and naturalness of the text determine the experience of the reader. It requires redundancy in the signal, which for our purposes indicates the need for the LM distribution over sequences to have high entropy. Therefore, constrained generation are more difficult to watermark because the entropy over the desired text distribution is much lower than in the case of open-ended generation. Consider a task like machine translation: given an input, there is a small set of acceptable outputs that convey similar meaning but differ greatly in terms of style, syntax, coverage, etc. This set is much smaller than open-ended generation but still larger and more varied than some low-entropy tasks like factual question-answering or even code-generation. Therefore, we posit and empirically observe with our proposed approach (\S~\ref{sec:main_results}), that these seemingly low-entropy constrained generation tasks have enough sequence-level entropy to enable imperceptible watermarking. Assuming max-length of sequences (denoted by r.v.~$\mathbf{y}$) is $T$, we can compute this entropy over $T$ random variables (tokens) under our model (parametrized by $\theta$) via chain rule for entropy: $H_\theta(\mathbf{y}) = \sum_{i=1}^T H_\theta(\mathbf{y_i}| \mathbf{y_{<i}})$ where crucially the conditional entropy involves summation over all the possible assignments of the prefix context $\mathbf{y_{<i}}$ i.e. $H(s \mid t) = -\sum_{s,t} p(s,t)~\log~p(s \mid t)$. This entropy is typically intractable to compute and is estimated \citep{kuhn2023semantic} via sampling. In a sharp contrast, token-level entropy that is often used to characterize uncertainty \citep{duan2023shifting} of language models is tractable to compute because instead of summing over all the possible prefixes, it commits to a single sampled prefix and computes the entropy over the possible tokens at the next step: $H_{\theta}(\mathbf{y_t}|y_{<t}) = - \sum_{w \in \mathcal{Y}}p(w\mid y_{<t})~\log~p(w\mid y_{<t})$. The sequence entropy that is typically computed using token-level entropy computation for a sequence $w = w_1,\ldots,w_T$ given by $H_{\theta}(w) = \sum_{i=1}^T H(w_i\mid w_{<i})$ is one-sequence approximation to the full intractable sequence-level entropy. Therefore, while tractable and attractive, common approaches of using token-level entropy to characterize sequence entropy tend to underestimate the true entropy over the sequences under the language model.

This observation has crucial ramifications for watermarking algorithms in low-entropy settings. The prevalent entropy-sensitive watermarking algorithms in prior work~\citep{lee-etal-2024-wrote, lu-etal-2024-entropy} only consider token-level entropy while embedding watermark in tokens during generation. For example, during tokenwise generation, \citet{lee-etal-2024-wrote} only chooses to mark the tokens that have high entropy under the language model. As we show in our experiments (\S~\ref{sec:main_results}), this approach does not perform well for constrained generation tasks. 
In Figure~\ref{fig:pareto}, we compare the token-level watermarking algorithm KGW~\citep{pmlr-v202-kirchenbauer23a} against two sequence-level watermarking algorithms, SemStamp~\citep{hou-etal-2024-semstamp} and our proposed approach SeqMark (\S~\ref{sec:seqmark}). By sweeping over hyperparameters for these algorithms, we obtain a pareto curve for each approach. It is clear that the the token-level watermarking approach is inferior to both sequence-level approaches -- we see a severe drop-off in watermark detection accuracy as the output quality increases for both the tasks.

As a token-level approach commits to a sequence during left-to-right generation, it ignores all other prefixes and potential paths that could lead to feasible outputs which drastically limits the entropy it can exploit for watermarking. Sequence-level semantic watermarking on the other hand considers sequence-level distribution via rejection sampling and is more amenable to exploit the sequence entropy afforded by constrained generation tasks.

\begin{figure*}[!t]
    \centering
    \includegraphics[width=0.95\textwidth]{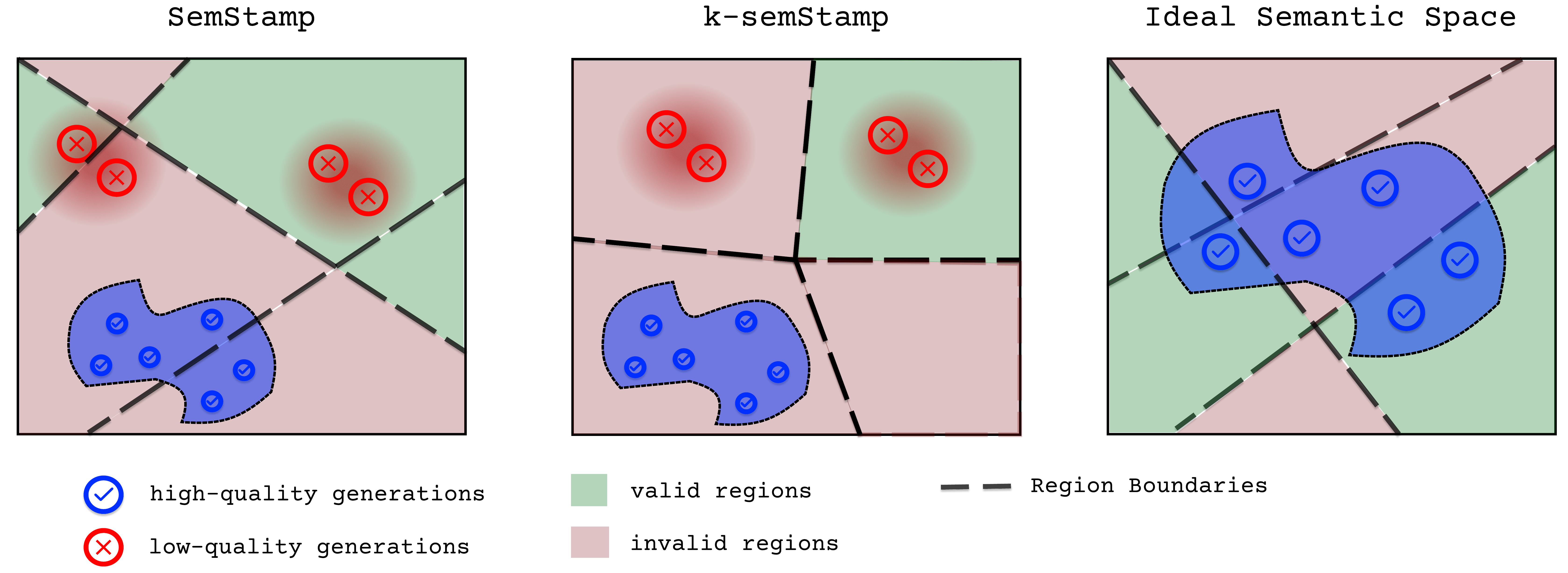}
    \caption{Semantic space illustration for different sequence watermarking algorithms. 
    Both SemStamp \textbf{(Left)} and $k$-SemStamp \textbf{(Middle)} suffer from region collapse. Ideally, we would like to isolate the high-quality output manifold and partition it \textbf{(Right)}.}
    \label{fig:motivation}
\end{figure*}

\begin{table}[!t]
\centering
\small 
\begin{tabular}{l|ccc}
\toprule 
\textbf{Approach} & \textbf{Translation} & \textbf{Summary} & \textbf{Open-ended}  \\
\midrule
\semstamp & 0.19 / 0.97 & 0.39 / 0.92  & 0.71 / 0.59 \\
\ksemstamp & 0.25 / 0.97 & 0.11 / 0.94 & 0.87 / 0.31 \\
\ours & 0.81 / 0.01 & 0.92 / 0.03 & 0.96 / 0.01 \\
\bottomrule
\end{tabular}
\caption{\textbf{Region entropy/average pairwise cosine similarity} across different tasks (columns) and watermarking approaches (rows). Higher region entropy means the points are more spread out.
} 
\label{tab:quantitative_motivation}
\vspace{-5mm}
\end{table}

\section{Region Collapse in Semantic Watermarking}
\label{sec:semstamp_insufficient}

We observe in Figure~\ref{fig:pareto} that the sequence-level algorithm SemStamp behaves better than the token-level algorithm KGW, but still is inferior to our approach described later. We identify a crucial property of sequence-level watermarking algorithms in prior work that render them unsuitable for constrained generation watermarking. We call this property \emph{region collapse} to refer to the effect that all viable outputs for a constrained generation task either randomly fall in the invalid red region or the valid green region. Consider the task of translation: If all the good translations of the prompt fall in the red region, then the sequence-level watermarking algorithms will either generate a bad translation from the green region and hurt the translation quality, or exhaust its resampling budget and output a translation from the red region, eventually degrading the watermark effectiveness. We describe the manifestation of region collapse in prior work on semantic watermarking algorithms: SemStamp~\citep{hou-etal-2024-semstamp} and $k$-SemStamp~\citep{hou-etal-2024-k}.

A well-trained language model induces a high probability on the correct responses for a constrained generation prompt and samples them more frequently.\footnote{We assume that highly probable generations from a well-trained LM are high-quality generations. While this might not always be the case, we believe this is a reasonable estimation for constrained generation tasks given the large($\sim$infinite) output sequence space, especially under well-trained language models. In addition, it is commonly observed that low-probability sequences tend to mostly be low-quality.} Also, because of semantic similarity of the high quality responses, the embeddings for these responses tend to have high similarity. An ideal sequence-level watermarking algorithm would produce red/green regions that separate these nearby responses to prevent region collapse as depicted in Figure~\ref{fig:motivation}. However, SemStamp's partitioning behaves in a manner \emph{opposite} to this desired behavior as it employs locality sensitive hashing (LSH) to determine the partitions. 

Under LSH \citep{charikar2002similarity, indykandmotwani1998}, given two vectors $\mathbf{x_i},\mathbf{x_j} \in \mathbb{R}^d$ with angle $\theta_{ij} \in [0, \pi]$, the probability that they lie in the same region is $( 1-\theta_{ij}/\pi)^d$.
LSH determines partitioning hyperplanes such that similar points share the same partition, thereby \emph{accelerating} region collapse. 
k-SemStamp, while shown to be more robust than SemStamp, unfortunately \emph{exacerbates} the region collapse issue for constrained generation. As described in \S \ref{sec:prelim},
this method explicitly focuses on assigning the same centroid/color to the semantically similar points with low embeddings distances, thus exacerbating region collapse.


In Table~\ref{tab:quantitative_motivation}, we empirically demonstrate the issue of region collapse on the tasks of translation and summarization. We sample $N=100$ high-quality (and high-probability) generations for each setting and report average pairwise cosine similarity and the estimated region entropy -- a quantity that characterizes how evenly the high-quality points are spread across the partitions. 
The region entropy is computed as $-\sum_{r \in \mathcal{R}} P(r) \log P(r)$, where $\mathcal{R}$ is the set of semantic regions. We estimate $P(r)$ using Monte Carlo sampling: $P(r) = 1/N \sum_{i=1}^N \mathbf{1}(\text{region}(x_i) = r)$.
While all three methods behave well for open-ended generation, for constrained generation tasks we observe low region entropy and high semantic similarity for SemStamp and $k$-SemStamp indicating region collapse, but high region entropy and low semantic similarity (explained below) for our SeqMark approach that explicitly ameliorates the issue of region collapse.

\section{SeqMark: Fixing Region Collapse}
\label{sec:seqmark}

\begin{figure*}[t]
    \centering
    \includegraphics[width=0.95\textwidth]{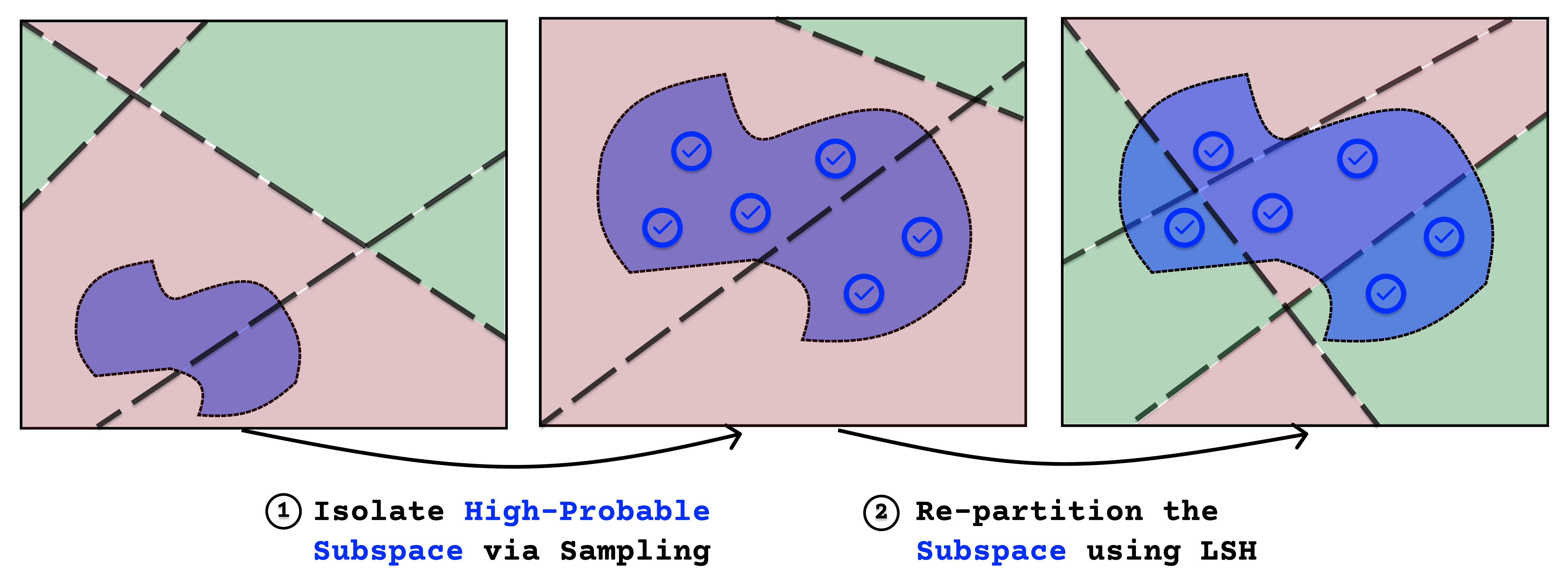}
    \caption{Depiction of SeqMark: Standard LSH in the first panel collapses the high-probable output subspace (\textcolor{blue}{blue}) to a small number of invalid regions (\textcolor{darkpink}{red}). We identify the high-probable subspace by sampling with low temperature; then we repartition by performing LSH on a \emph{transformed} subspace designed to evenly spread it across multiple regions.}
    \vspace{-5mm}
    \label{fig:approach}
\end{figure*}

As mentioned above, we propose SeqMark, a sequence-level watermarking approach that addresses region collapse. 
Similar to other semantic watermark approaches \citep{hou-etal-2024-semstamp, hou-etal-2024-k}, SeqMark pseudorandomly partitions the representation space into accept/reject regions and performs rejection sampling for watermarking.
However, it differs significantly in how it operationalizes the partitioning procedure. As shown in Figure~\ref{fig:approach}, we first focus on isolating the subspace manifold containing high quality responses for the constrained generation prompt. We estimate this manifold by sampling highly-likely generations using low-temperature sampling.
Once isolated, we partition this manifold such that the high quality (highly probable) points are evenly distributed among the regions. To perform this partitioning, we still use LSH, but crucially we \emph{transform the isolated manifold} such that its members (high-quality responses) are distant from one another, causing LSH to evenly spread them across the random partitions.
Concretely, for an input prompt $p$ (that includes $t-1$ previously generated sentences), we generate the response ($t$-th sentence $s_t$) as follows: 
We first sample $n$ high likelihood responses (with embeddings $\mathbf{c_i}$) under low temperature to estimate high-quality subspace $\mathcal{C} = \{\mathbf{c_1}, \mathbf{c_2},...,\mathbf{c_n}\}$. 
Let $f(\cdot; \mathcal{C})$ be the function that transforms its members $\mathbf{c_i}$ into a representation $\mathbf{u_i}$: $\mathbf{u_i} = f(\mathbf{c_i} ; \mathcal{C})$, which would be used to assign an LSH partition to $\mathbf{c_i}$. To prevent region collapse, we aim to  minimize the pairwise cosine similarity between the transformed members of $\mathcal{C}$ while preserving their relative proximity.  
This is difficult to estimate in general and could be approximated by learning such a function via a neural network. We opt for a much simpler choice for $f$: subtracting the sample mean from each member of $\mathcal{C}$:
$f(\mathbf{c_i}; \mathcal{C}) = \mathbf{c_i} - \frac{1}{n} \sum_{j=1}^n \mathbf{c_j}$.\footnote{The full generation and detection algorithms are given in Algorithm \ref{alg:generate} and Algorithm \ref{alg:detect}, respectively.}
We empirically observe that our mean-centering approach increases the distance between high-quality points in the desired manner (Table~\ref{tab:quantitative_motivation}). Finally, Theorem~1 below states that watermark detection accuracy improves with this mean-centering transformation.
\vspace{-2.5mm}
\paragraph{Theorem 1} \label{theorem:1} Given $n$ vectors $\mathbf{c_1},...,\mathbf{c_n}$ such that  
$\mathbf{c_i} = \mu + \epsilon_i$ and there exists $\delta \in [0,1]$ such that $||\epsilon_i|| \leq \delta||\mu||$. Mean-centering transformation results in vectors $\mathbf{u_i} = \mathbf{c_i} - \frac{1}{n}\sum_i \mathbf{c_i} = \epsilon_i$. If the following condition holds for a fixed pair $i,j$:
$$cos \theta^{(\mathbf{u})}_{ij} = \frac{\epsilon_i \cdot \epsilon_j}{||\epsilon_i||||\epsilon_j||} \leq \frac{1-2\delta - \delta^2}{1+2\delta+\delta^2}$$
then the watermark detection accuracy obtained using an LSH partition of
$\mathbf{u}_1,\dots,\mathbf{u}_n$ improves upon that obtained using an
LSH partition of $\mathbf{c}_1,\dots,\mathbf{c}_n$.
\vspace{-2.2mm}
\paragraph{Remark} The constant $\delta$ summarizes vector similarity. Intuitively, the mean term $\mu$ of extremely similar vectors would dominate the residual term $\epsilon$, thus $\delta << 1$ and $\frac{1-2\delta - \delta^2}{1+2\delta+\delta^2}$ would be close to 1. This gives a rather relaxed bound for residuals above, and almost guarantee that the transformation would result in increasing detection accuracy. The full proof and discussion are presented in \S \ref{app:performance_proof}.


\section{Experimental Setup}

\label{sec:main_experiment}
\noindent \textbf{Tasks and Datasets:} We evaluate our approach on several constrained text generation settings: sentence translation with WMT19 German-English dataset \citep{wmt19translate}, paragraph translation with WMT23 German-English \citep{kocmi-etal-2023-findings}, abstractive summarization with XSum \citep{narayan-etal-2018-dont}, and code generation with MBPP \citep{austin2021programsynthesislargelanguage}. For unconstrained open-ended generation, we follow previous work and complete sentences from C4 RealNews dataset \citep{c4realnews}. Table \ref{tab:datasets} presents tasks and datasets details.

\noindent \textbf{Language Models and Embeddings:} For sentence translation, we use ALMA-7B \citep{alma7b} and LaBSE \citep{feng-etal-2022-language} as the primary language model and sentence encoder, respectively. For paragraph translation, we use Gemma-2-4B-it \citep{gemma2_2024} due to its long context window. For code generation, we use CodeLlama-7B \citep{roziere2024codellamaopenfoundation} and Gemma embeddings \citep{vera2025embeddinggemmapowerfullightweighttext}. For other settings, we use Llama-2-7B-Chat \citep{touvron2023llama2openfoundation} and SBERT encoder \citep{reimers-2019-sentence-bert}.\footnote{Further experiment with sentence encoders is in \S \ref{app:encoder}.}

\noindent \textbf{Baselines:} We compare our approach against token-level watermarking approaches KGW~\citep{pmlr-v202-kirchenbauer23a}, and SWEET~\citep{lee-etal-2024-wrote}, and sequence-level watermarking approaches SemStamp~\citep{hou-etal-2024-semstamp}, and $k$-SemStamp~\citep{hou-etal-2024-k}. 
Given different sets of hyperparameters for each algorithm, we first sweep for the best hyperparameters over 100 samples to find the most promising configurations, then run them on the full evaluation sets (\S \ref{app:hyperparams}).

\noindent \textbf{Evaluation:} Two important dimensions for evaluating watermark algorithms are \textit{text quality} and \textit{watermark detectablity}. Text quality metrics often depend on the specific tasks and are thus described in the corresponding sections below. For watermark detection, previous works utilizing statistical hypothesis testing often report classification metrics such as AUROC and True Positive Rate (at a specific False Positive Rate, e.g. FPR=1\%). While we also report these metrics for longer generation tasks (paragraph translation, open-ended generation), for sentence translation, summarization, and code generation these metrics are ill-suited since the hypothesis test is often over a single sequence.  
Therefore, we treat each query as a binary classification problem: either it was detected as a watermark or not. We then compute precision, recall, and \fscore{} on the detection predictions.\footnote{Alternatively, one can aggregate the sequences over multiple examples in the evaluation set for AUROC computation. We experimented with this aggregated AUROC metric in \S \ref{app:full} and found that the results are consistent with our conclusions.}

Finally, we report detection results for two types of negative examples: a) human (\textbf{h}) to evaluate differentiation between watermarked LM and human generated text, and b) non-watermarked LM (\textbf{nw}), to evaluate differentiation between watermarked generation and non-watermarked generation from the same language model.

\section{Results and Analysis}
\label{sec:main_results}

\subsection{Main Results}

\begin{table*}[t]
\centering
\small 
\setlength{\tabcolsep}{3pt} 
\renewcommand{\arraystretch}{1.15} 
\resizebox{\textwidth}{!}{%
\begin{tabular}{lcccccc}
\toprule
& \multicolumn{3}{c}{\textbf{Translation (WMT19 De-En)}} 
& \multicolumn{3}{c}{\textbf{Summarization (XSum)}} \\
\cmidrule(lr){2-4} \cmidrule(lr){5-7} 
& \textbf{COMET} $\uparrow$ & \textbf{P / R / \fscore (h)} &  \textbf{P / R / \fscore (nw)} & \textbf{R-L} $\uparrow$ / \textbf{ COMET} $\uparrow$ & \textbf{P / R / \fscore (h)} & \textbf{P / R / \fscore (nw)} \\
\midrule
\textit{No Watermark} & \textit{87.4} & - & - & \textit{20.5 / 69.0} & - & - \\
\kgw        & \underline{87.4} & 57.2 / 36.6 / 45.7 & 58.8 / 40.0 / 47.6 & 20.1 / \underline{68.8} & \underline{85.2} / 30.1 / 44.9 & \underline{79.8} / 30.4 / 44.0 \\
\sweet & 87.2 & 57.5 / 30.0 / 39.5 & 58.1 / 30.1 / 39.6 &  18.5 / 68.7 & 52.4 / 49.9 / 51.1 & 53.0 / 40.0 / 45.6 \\
\semstamp    & \underline{87.4} & 59.5 / \underline{73.7} / \underline{65.9} & 58.7 / \underline{74.0} / \underline{65.5} & 20.0 / 68.7 & 70.6 / \underline{52.8} / \underline{60.4} & 67.8 / \underline{59.0} / \underline{63.1} \\
\ksemstamp  & \textbf{87.5} & \underline{63.3} / 33.0 / 43.4 & \underline{62.2} / 33.0 / 43.1 & \underline{20.7} / \textbf{68.9} & 54.2 / 22.0 / 31.3 & 61.6 / 22.0 / 32.4 \\
\ours & 87.1 & \textbf{76.9 / 77.3 / 77.1} & \textbf{75.5 / 83.0 / 79.0} & \textbf{21.6} / 68.5 & \textbf{81.3 / 100 / 89.7} & \textbf{85.3 / 85.3 / 85.3} \\
\bottomrule
\end{tabular}
}
\caption{Watermarking results for sentence translation and summarization. Best results are \textbf{bold} and second best are \underline{underlined}. $\uparrow$ denotes the higher the better. For detection, \textbf{(h)} and \textbf{(nw)} denote the negative examples from human and non-watermarked LM, respectively. \ours substantially improves detection while maintaining competitive text quality.}
\vspace{-2mm}
\label{tab:constrained}
\end{table*}

\begin{table*}[t]
\centering
\small 
\setlength{\tabcolsep}{3pt} 
\renewcommand{\arraystretch}{1.15} 
\resizebox{\textwidth}{!}{%
\begin{tabular}{lcccccc}
\toprule
& \multicolumn{3}{c}{\textbf{Code Generation (MBPP)}} 
& \multicolumn{3}{c}{\textbf{Open-ended Generation (C4)}} \\
\cmidrule(lr){2-4} \cmidrule(lr){5-7} 
& \textbf{pass@1} $\uparrow$ & \textbf{P / R / \fscore (h)} &  \textbf{P / R / \fscore (nw)} & \tiny \textbf{PPL} $\downarrow$ & \tiny \textbf{AUROC / TP@1 / TP@5 (h)} & \tiny \textbf{AUROC / TP@1 / TP@5 (nw)} \\
\midrule
\textit{No Watermark} & \textit{33.8} & - & - & \textit{3.4} & - & - \\
\kgw        &  15.0 & \textbf{87.5} / 31.0 / 45.7 & \textbf{86.5} / 33.0 / 45.9 & 3.6 & \underline{99.0} / 92.0 / 95.0 & \textbf{99.9 / 99.0 / 100} \\
\sweet &  33.2 & \underline{79.5} / 62.3 / \underline{70.2}  & \underline{80.5} / 63.2 / \underline{71.6} & 3.6 & 97.6 / \underline{93.0} / \textbf{97.0} & 97.3 / 93.0 / 97.0 \\
\semstamp    & \textbf{34.2} & 66.9 / \underline{71.8} / 69.2 & 66.8 / 72.0 / 69.1 & 3.6 & 98.7 / 90.9 / \textbf{97.0} & \underline{99.0} / 85.9 / \underline{98.0} \\
\ksemstamp & \underline{34.0}	& 59.9 / 71.7 / 65.2 & 58.7 / \underline{74.0} / 65.5 & 3.6 & 94.8 / 80.0 / 87.0 & 94.8 / 80.0 / 87.0 \\
\ours &  33.6 & 76.0 / \textbf{86.0} / \textbf{80.7} & 76.3 / \textbf{87.0} / \textbf{81.3} & \textbf{3.4} & \textbf{99.3 / 94.0 / 97.0} & \underline{99.0} / \underline{94.0} / 96.0 \\
\bottomrule
\end{tabular}
}
\caption{Code generation with MBPP and open-ended generation with C4 RealNews}
\vspace{-5mm}
\label{tab:code_and_c4}
\end{table*}

\noindent \textbf{Sentence translation, summarization, and code generation:} For translation evaluation, we use COMET, a neural, semantic-based metric \citep{rei-etal-2020-comet}. Since COMET is not specifically trained for summarization, we include ROUGE-L \citep{lin-2004-rouge} for summarization results. For code generation, we use pass@1 metric. We observe in Table \ref{tab:constrained} that SeqMark substantially improves detection results while maintaining competitive text quality for sentence translation and summarization. This result is in congruent with Table \ref{tab:quantitative_motivation}, where the high cluster entropy score and low cosine similarity indicates that SeqMark correctly defines the high-quality output space and partitions it more uniformly for more effective watermarking.  
While SemStamp improves over token-level approaches, we confirm our hypothesis that $k$-SemStamp exacerbates region collapse by observing that it performs worse than SemStamp. For code generation, SeqMark outperforms previous best baseline SWEET (Table \ref{tab:code_and_c4}). \ Finally, the trends are similar for both the types of negatives, human and non-watermarked LM, indicating that SeqMark successfully differentiates against texts from other sources.

\noindent \textbf{Open-ended generation:} Similar to previous work, we evaluate watermarking algorithms on open-ended generations using C4 RealNews (Table~\ref{tab:code_and_c4}). All watermarking algorithms perform well in this high-entropy setting, with SeqMark performing on par with other baselines without causing significant degradation in perplexity.

\begin{wraptable}{r}{0.45\textwidth}
\centering
\small 
\setlength{\tabcolsep}{1.5pt} 
\renewcommand{\arraystretch}{1.1} 
\begin{tabular}{lcc}
\toprule
\tiny \textbf{Approach} & \tiny \textbf{BLEU}/ \textbf{COMET} & \tiny \textbf{AUROC / TP@1 / TP@5 (h)}\\
\midrule
\textit{No Watermark} & \textit{39.2 / 87.1} & - \\
\kgw        & 40.7 / \textbf{87.7} & 72.1 / 11.0 / 25.0 \\
\sweet & 39.2 / 87.4 & 60.6 / 10.0 / 16.2\\
\semstamp    &  \textbf{42.8} / 87.5 & \underline{84.0} / \underline{30.0} / \underline{40.0} \\
\ksemstamp & \underline{42.2} / \underline{87.6} & 53.5 / 1.0 / 6.0 \\
\ours &  39.8 / \textbf{87.7} & \textbf{100 / 100 / 100} \\
\bottomrule
\end{tabular}
\caption{Paragraph translation with WMT23}
\vspace{-2mm}
\label{tab:wmt23}
\end{wraptable}

\noindent \textbf{Paragraph translation:} We also evaluate our approach on paragraph translation, where the LM is tasked with translating multiple sentences. For this task, we concatenate 8-10 sentences from the same article in WMT23 dataset such that the input prompt fits within the context window of Gemma-2-4B-it. Since COMET is not specifically trained for paragraph evaluation, we also report BLEU \citep{papineni-etal-2002-bleu}. Results in Table \ref{tab:wmt23} shows a similar trend as Table \ref{tab:constrained}: SeqMark has significantly higher detection score than KGW and SemStamp, while maintaining similar text quality.


\subsection{Effect of Transformations $f$}


\begin{wraptable}{r}{0.5\textwidth}
\centering
\small 
\setlength{\tabcolsep}{1.5pt} 
\renewcommand{\arraystretch}{1.15} 
\begin{tabular}{lcc}
\toprule
\tiny \textbf{Approach} & \tiny \textbf{COMET $\uparrow$} & \tiny \textbf{P / R / \fscore (h)} \\
\midrule
\kgw        & \textbf{85.5}  & 67.2 / 37.0 / 47.8 \\
\semstamp    & 85.2 & 58.7 / 74.0 / 65.5 \\
Sample mean & 85.0 & \textbf{75.9} / \underline{85.0} / \underline{80.2} \\
\midrule
Random embedding & \underline{85.4} & 51.3 / 19.0 / 27.7\\
Sample closest-to-mean & \underline{85.4} & \underline{75.0} / \textbf{87.0} / \textbf{80.6}  \\
Single sample & 85.3 & 71.1 / 59.0 / 64.5 \\
Source embedding & 84.9 & 67.9 / 53.0 / 59.6 \\
Target embedding & \underline{85.4} & 64.8 / 59.0 / 61.8 \\ 
\bottomrule
\end{tabular}
\vspace{-1mm}
\caption{Different transformations $f$ on WMT19}
\label{tab:transformation_ablation}
\end{wraptable}

The goal of the transformation $f$ in SeqMark is to reduce cosine similarity between high-quality generations. We propose a transformation that modifies sentence embedding by subtracting the sample mean i.e. $f(\mathbf{c_i}; C) = \mathbf{c_i} - \mathbf{z}$, where $\mathbf{z} = 1/n\sum_{\mathbf{c} \in C} c$. In Table~\ref{tab:transformation_ablation}, we investigate other transformations obtained by subtracting vectors $\mathbf{z}$ other than the mean from the embeddings: random embedding, the sample closest to the mean, a single sample embedding $\mathbf{c}$, the source sentence embedding, and the ground truth target translation embedding. 
We observe that all transformations perform better than subtracting a random embedding, indicating the need for preservation of relative relationships among the samples. Importantly, aggregated representations such as sample mean and point closest-to-the-mean perform the best, while single point embedding alternatives do not outperform SemStamp baseline. 

\vspace{2mm}

\subsection{Computational Cost and Fast-SeqMark} 



\begin{wrapfigure}{r}{0.45\textwidth}
    \centering
    \includegraphics[width=\linewidth]{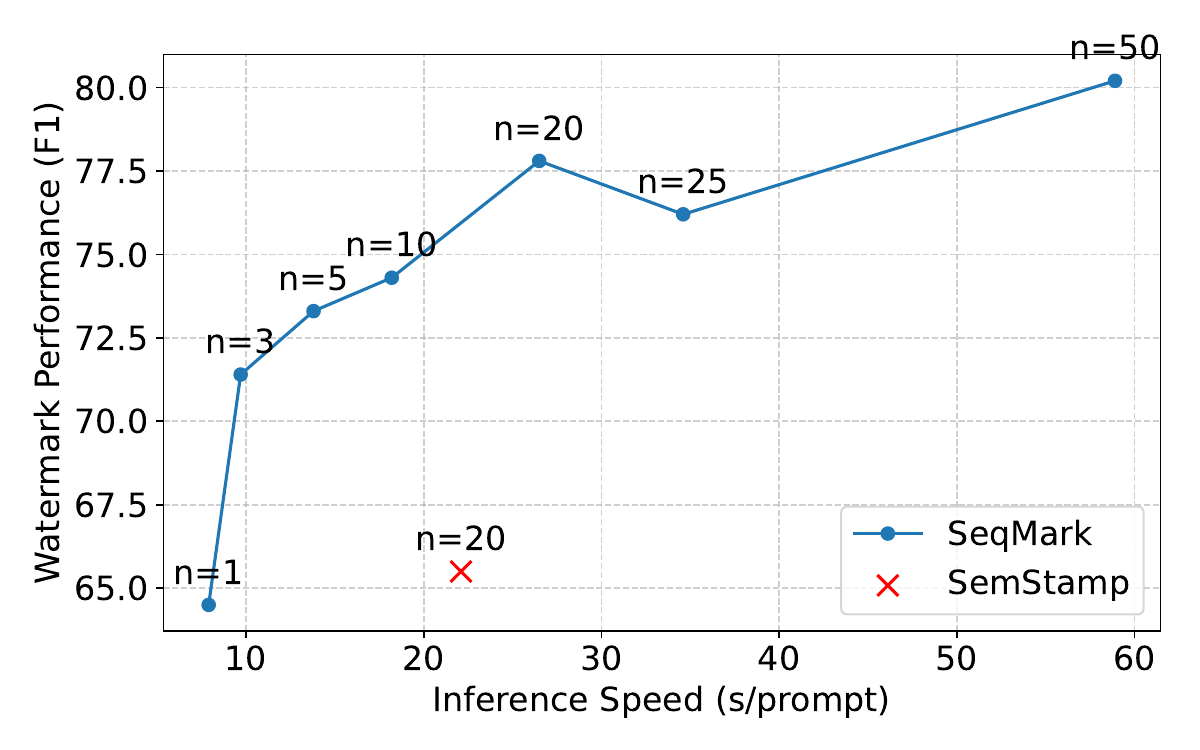}
    \caption{Performance-speed trade-off for varying $n$. For text quality, COMET scores range from 84.9-85.3, well within the score of SemStamp (85.2)}
    \vspace{-3mm}
    \label{fig:varying_n}
\end{wrapfigure}

\paragraph{Varying number of samples to compute mean $\mathbf{z}$} In all our experiments, we use $n=50$ as the number of sampled sequences to compute the mean embedding $\mathbf{z}$. To measure the effect of $n$ on the inference efficiency vs. performance tradeoffs of 
SeqMark, we vary $n$ across different values, using WMT19 as evaluation data. We benchmark the results against the inference speed of sequence-level algorithm SemStamp, using $n=20$ samples. Inference speed is measured by seconds per input prompt. The experiment is performed on a single NVIDIA A40 GPU. Result in Figure~\ref{fig:varying_n} shows that as both inference speed and watermark performance increase as we increase $n$. However, watermark performance plateaus for values of $n$ above 20, while inference speed for $n=10$ and $n=20$ are comparable to SemStamp. Thus, a value of $n$ between 10-20 samples presents a reasonable choice for balancing between watermark performance and inference speed.

\vspace{-3mm}

\begin{wraptable}{r}{0.43\textwidth}
\centering
\small 
\setlength{\tabcolsep}{3pt} 
\renewcommand{\arraystretch}{1.15} 
\vspace*{-\baselineskip}
\begin{tabular}{lcc}
\toprule
\tiny \textbf{Approach} & \tiny \textbf{COMET $\uparrow$} & \tiny \textbf{P / R / \fscore (h)} \\
\midrule
\kgw        & \textbf{85.5}  & 67.2 / 37.0 / 47.8 \\
\semstamp    & \underline{85.2} & 58.7 / 74.0 / 65.5 \\
\ours & 85.0 & \textbf{75.9 / 85.0 / 80.2} \\
\texttt{Fast-}\ours & 84.4  & \underline{68.1} / \underline{75.0} / \underline{71.4}\\
\bottomrule
\end{tabular}
\caption{\texttt{Fast-}\ours on WMT19}
\vspace{-5mm}
\label{tab:fast_correct}
\end{wraptable}

\paragraph{Fast-SeqMark} As mentioned above, SeqMark samples high-quality generations to estimate the sample mean for the transformation.
Consequently during detection, computing the mean embedding via sampling requires computation and access to the language model for detection. To avoid this step altogether, we propose a lightweight solution where a neural network is trained to map the input prompt to the sample mean embedding $\mathbf{z}$ to subtract. We experiment with this Fast-SeqMark solution on the WMT19 German-English dataset. For training data generation, we use ALMA-7B to generate 50 samples each for 100,000 input prompts with a fixed temperature of 1.2.
We then finetune the LABSE encoder on this dataset for mean prediction. 
Table \ref{tab:fast_correct} shows that when compared to using the actual mean embeddings, Fast-SeqMark results in a drop in detection score, yet still outperforms KGW and SemStamp.



\section{Related Work}


\paragraph{Language Model Watermarking} Since its introduction, LM Watermarking has become an important technique to combat LLM-related security concerns \citep{Suvra2023ASO, brookingswatermark}. Most algorithms propose adding imperceptible statistical signals to different stages of text generation, including logit generation \citep{pmlr-v202-kirchenbauer23a, Zhao2023ProvableRW, hu2023unbiased, liu2024semanticinvariantrobustwatermark}, token sampling \citep{christ2024undetectable, kuditipudi2023robust, mao-etal-2025-watermarking}, or the model weights \citep{sun2023codemark, gu2023learnability}. Orthogonally, several work propose sequence-level watermarking \citep{hou-etal-2024-semstamp, hou-etal-2024-k} that involves sampling full sentences, encodes them into a semantic space, and selects the generations within the valid semantic regions. 
Overall, the majority of work in LM watermarking concerns with open-ended text generation which affords high-entropy thus enabling highly effective watermarking \citep{ajith-etal-2024-downstream}, leaving watermarking for constrained text generation tasks such as machine translation and summarization remains underexplored, despite these use cases constituting a non-trivial proportion of LLM usage.


\paragraph{Watermarking Low-Entropy Sequences} Prior work on watermarking low-entropy sequences has mostly focused on code generation: \citet{lee-etal-2024-wrote} extends \citet{pmlr-v202-kirchenbauer23a} by selecting and watermarking high-entropy tokens, \citet{lu-etal-2024-entropy} modifies the detection algorithm to include token entropy as weights in the final score, and \citet{gu2025invisibleentropysafeefficient} effectively reduces watermarking efficiency without major loss in performance. 
Similar to us, \cite{cai2025entropyguidedwatermarkingllmstesttime} and \cite{tsur2025heavywatersimplexwaterdistortionfreellm} also addressed the trade-offs in watermarking low-entropy outputs; however, their work only consider token-level entropy. For translation, \citet{takezawa2025necessarysufficientwatermarklarge} modifies KGW by considering the minimal number of watermarked tokens needed; nonetheless, their approach is not imperceptible-- it is visually easy to detect the watermarked tokens (\S \ref{app:imperceptible}). Orthogonal to all previous work focusing on extending token-level watermarking, our work focuses on the strengths and weaknesses of sequence-level watermarking, such as the region collapse issue for watermarking low-entropy texts. Interestingly, we observe that sequence-level watermarking outperforms token-level algorithms in general, possibly due to underestimation of uncertainty by token-level algorithms \citep{kuhn2023semantic}.

\vspace{-3mm}

\section{Conclusion}

\vspace{-3mm}

In this work, we demonstrate the inadequacy of current watermarking algorithms for constrained generation tasks. We posit that all token-level algorithms perform poorly because they fail to utilize the semantic entropy induced by the LMs on these tasks. While sequence level watermarking algorithms are a better paradigm, we identify a different issue of region collapse in the operationalization of existing semantic watermarking algorithms, resulting in poor performance at watermarking constrained generation tasks.
To overcome these limitations, we propose SeqMark, a sequence-level watermarking algorithm that ameliorates the region collapse issue by carefully partitioning the space for watermarking so as to ensure an even spread of high-quality outputs among the partitions. This results in reliably detectable watermarks while generating high-quality outputs for low-entropy tasks.

\vspace{-3mm}

\section*{Limitations}

\vspace{-3mm}

During detection, SeqMark requires access to the language model as well as precise reproduction of the sampled generations to compute mean embeddings. Despite the fact that these are not ideal, we present Fast-SeqMark as a viable alternative that alleviates these requirements. In addition, prior work~\citep{lee-etal-2024-wrote, lu-etal-2024-entropy} also requires access to the LM during detection for computing token-level entropy. This access is readily available for open-weights models and could also be requested for realistic auditing purposes.


Another limitation is that, unlike SemStamp / $k$-SemStamp, SeqMark might be susceptible to paraphrasing attacks. By design, SeqMark spreads similar generations more evenly in the semantic space; thus a strong paraphraser can potentially evade the watermark by transforming the generation into another semantic region. However, in our preliminary paraphrase attack experiment (\S \ref{app:paraphrase}), we observe that the paraphrasing outputs often have low text-quality, indicating that current paraphrasing models are not strong enough to maintain text quality, especially for low-entropy tasks like translations. 

Like other LM watermarking technologies, SeqMark is designed to help enhance safe usages of language models. Nonetheless, it also carries potential risks that come with misusing these technologies. For example, it is susceptible to spoofing attack, where a malicious actor takes advantage of watermark robustness by embedding the watermark into unethical texts, which can then be misattributed to specific language model providers. Therefore, we recommend users exercise caution to avoid exposing important details such as the LSH hash function key.

\bibliographystyle{colm2026_conference}
\bibliography{colm2026_conference}

@misc{tsur2025heavywatersimplexwaterdistortionfreellm,
      title={HeavyWater and SimplexWater: Distortion-Free LLM Watermarks for Low-Entropy Next-Token Predictions}, 
      author={Dor Tsur and Carol Xuan Long and Claudio Mayrink Verdun and Hsiang Hsu and Chen-Fu Chen and Haim Permuter and Sajani Vithana and Flavio P. Calmon},
      year={2025},
      eprint={2506.06409},
      archivePrefix={arXiv},
      primaryClass={cs.CR},
      url={https://arxiv.org/abs/2506.06409}, 
}

@inproceedings{mao-etal-2025-watermarking,
    title = "Watermarking Large Language Models: An Unbiased and Low-risk Method",
    author = "Mao, Minjia  and
      Wei, Dongjun  and
      Chen, Zeyu  and
      Fang, Xiao  and
      Chau, Michael",
    editor = "Che, Wanxiang  and
      Nabende, Joyce  and
      Shutova, Ekaterina  and
      Pilehvar, Mohammad Taher",
    booktitle = "Proceedings of the 63rd Annual Meeting of the Association for Computational Linguistics (Volume 1: Long Papers)",
    month = jul,
    year = "2025",
    address = "Vienna, Austria",
    publisher = "Association for Computational Linguistics",
    url = "https://aclanthology.org/2025.acl-long.391/",
    doi = "10.18653/v1/2025.acl-long.391",
    pages = "7939--7960",
    ISBN = "979-8-89176-251-0"
}

@misc{cai2025entropyguidedwatermarkingllmstesttime,
      title={Entropy-Guided Watermarking for LLMs: A Test-Time Framework for Robust and Traceable Text Generation}, 
      author={Shizhan Cai and Liang Ding and Dacheng Tao},
      year={2025},
      eprint={2504.12108},
      archivePrefix={arXiv},
      primaryClass={cs.CL},
      url={https://arxiv.org/abs/2504.12108}, 
}

@misc{roziere2024codellamaopenfoundation,
      title={Code Llama: Open Foundation Models for Code}, 
      author={Baptiste Rozière and Jonas Gehring and Fabian Gloeckle and Sten Sootla and Itai Gat and Xiaoqing Ellen Tan and Yossi Adi and Jingyu Liu and Romain Sauvestre and Tal Remez and Jérémy Rapin and Artyom Kozhevnikov and Ivan Evtimov and Joanna Bitton and Manish Bhatt and Cristian Canton Ferrer and Aaron Grattafiori and Wenhan Xiong and Alexandre Défossez and Jade Copet and Faisal Azhar and Hugo Touvron and Louis Martin and Nicolas Usunier and Thomas Scialom and Gabriel Synnaeve},
      year={2024},
      eprint={2308.12950},
      archivePrefix={arXiv},
      primaryClass={cs.CL},
      url={https://arxiv.org/abs/2308.12950}, 
}

@misc{vera2025embeddinggemmapowerfullightweighttext,
      title={EmbeddingGemma: Powerful and Lightweight Text Representations}, 
      author={Henrique Schechter Vera and Sahil Dua and Biao Zhang and Daniel Salz and Ryan Mullins and Sindhu Raghuram Panyam and Sara Smoot and Iftekhar Naim and Joe Zou and Feiyang Chen and Daniel Cer and Alice Lisak and Min Choi and Lucas Gonzalez and Omar Sanseviero and Glenn Cameron and Ian Ballantyne and Kat Black and Kaifeng Chen and Weiyi Wang and Zhe Li and Gus Martins and Jinhyuk Lee and Mark Sherwood and Juyeong Ji and Renjie Wu and Jingxiao Zheng and Jyotinder Singh and Abheesht Sharma and Divyashree Sreepathihalli and Aashi Jain and Adham Elarabawy and AJ Co and Andreas Doumanoglou and Babak Samari and Ben Hora and Brian Potetz and Dahun Kim and Enrique Alfonseca and Fedor Moiseev and Feng Han and Frank Palma Gomez and Gustavo Hernández Ábrego and Hesen Zhang and Hui Hui and Jay Han and Karan Gill and Ke Chen and Koert Chen and Madhuri Shanbhogue and Michael Boratko and Paul Suganthan and Sai Meher Karthik Duddu and Sandeep Mariserla and Setareh Ariafar and Shanfeng Zhang and Shijie Zhang and Simon Baumgartner and Sonam Goenka and Steve Qiu and Tanmaya Dabral and Trevor Walker and Vikram Rao and Waleed Khawaja and Wenlei Zhou and Xiaoqi Ren and Ye Xia and Yichang Chen and Yi-Ting Chen and Zhe Dong and Zhongli Ding and Francesco Visin and Gaël Liu and Jiageng Zhang and Kathleen Kenealy and Michelle Casbon and Ravin Kumar and Thomas Mesnard and Zach Gleicher and Cormac Brick and Olivier Lacombe and Adam Roberts and Qin Yin and Yunhsuan Sung and Raphael Hoffmann and Tris Warkentin and Armand Joulin and Tom Duerig and Mojtaba Seyedhosseini},
      year={2025},
      eprint={2509.20354},
      archivePrefix={arXiv},
      primaryClass={cs.CL},
      url={https://arxiv.org/abs/2509.20354}, 
}

@misc{austin2021programsynthesislargelanguage,
      title={Program Synthesis with Large Language Models}, 
      author={Jacob Austin and Augustus Odena and Maxwell Nye and Maarten Bosma and Henryk Michalewski and David Dohan and Ellen Jiang and Carrie Cai and Michael Terry and Quoc Le and Charles Sutton},
      year={2021},
      eprint={2108.07732},
      archivePrefix={arXiv},
      primaryClass={cs.PL},
      url={https://arxiv.org/abs/2108.07732}, 
}

@misc{morris2024contextualdocumentembeddings,
      title={Contextual Document Embeddings}, 
      author={John X. Morris and Alexander M. Rush},
      year={2024},
      eprint={2410.02525},
      archivePrefix={arXiv},
      primaryClass={cs.CL},
      url={https://arxiv.org/abs/2410.02525}, 
}

@inproceedings{motwani2024secret,
 author = {Motwani, Sumeet Ramesh and Baranchuk, Mikhail and Strohmeier, Martin and Bolina, Vijay and Torr, Philip H.S. and Hammond, Lewis and de Witt, Christian Schroeder},
 booktitle = {Advances in Neural Information Processing Systems},
 doi = {10.52202/079017-2336},
 editor = {A. Globerson and L. Mackey and D. Belgrave and A. Fan and U. Paquet and J. Tomczak and C. Zhang},
 pages = {73439--73486},
 publisher = {Curran Associates, Inc.},
 title = {Secret Collusion among AI Agents: Multi-Agent Deception via Steganography},
 url = {https://proceedings.neurips.cc/paper_files/paper/2024/file/861f7dad098aec1c3560fb7add468d41-Paper-Conference.pdf},
 volume = {37},
 year = {2024}
}

@article{augenstein2024factuality,
  title={Factuality challenges in the era of large language models and opportunities for fact-checking},
  author={Augenstein, Isabelle and Baldwin, Timothy and Cha, Meeyoung and Chakraborty, Tanmoy and Ciampaglia, Giovanni Luca and Corney, David and DiResta, Renee and Ferrara, Emilio and Hale, Scott and Halevy, Alon and others},
  journal={Nature Machine Intelligence},
  volume={6},
  number={8},
  pages={852--863},
  year={2024},
  publisher={Nature Publishing Group UK London}
}

@article{duan2023shifting,
  title={Shifting attention to relevance: Towards the predictive uncertainty quantification of free-form large language models},
  author={Duan, Jinhao and Cheng, Hao and Wang, Shiqi and Zavalny, Alex and Wang, Chenan and Xu, Renjing and Kailkhura, Bhavya and Xu, Kaidi},
  journal={arXiv preprint arXiv:2307.01379},
  year={2023}
}

@article{kuhn2023semantic,
  title={Semantic uncertainty: Linguistic invariances for uncertainty estimation in natural language generation},
  author={Kuhn, Lorenz and Gal, Yarin and Farquhar, Sebastian},
  journal={arXiv preprint arXiv:2302.09664},
  year={2023}
}

@article {Gravel2023.03.16.23286914,
	author = {Gravel, Jocelyn and D{\textquoteright}Amours-Gravel, Madeleine and Osmanlliu, Esli},
	title = {Learning to fake it: limited responses and fabricated references provided by ChatGPT for medical questions},
	elocation-id = {2023.03.16.23286914},
	year = {2023},
	doi = {10.1101/2023.03.16.23286914},
	publisher = {Cold Spring Harbor Laboratory Press},
	URL = {https://www.medrxiv.org/content/early/2023/03/24/2023.03.16.23286914},
	eprint = {https://www.medrxiv.org/content/early/2023/03/24/2023.03.16.23286914.full.pdf},
	journal = {medRxiv}
}

@inproceedings{copyright,
author = {Zhong, Haonan and Chang, Jiamin and Yang, Ziyue and Wu, Tingmin and Mahawaga Arachchige, Pathum Chamikara and Pathmabandu, Chehara and Xue, Minhui},
title = {Copyright Protection and Accountability of Generative AI: Attack, Watermarking and Attribution},
year = {2023},
isbn = {9781450394192},
publisher = {Association for Computing Machinery},
address = {New York, NY, USA},
url = {https://doi.org/10.1145/3543873.3587321},
doi = {10.1145/3543873.3587321},
booktitle = {Companion Proceedings of the ACM Web Conference 2023},
pages = {94–98},
numpages = {5},
location = {Austin, TX, USA},
series = {WWW '23 Companion}
}

@article{Info2023ChatGPTIH,
  title={ChatGPT in higher education: Considerations for academic integrity and student learning},
  author={Miriam Sullivan and Andrew Kelly and Paul Mclaughlan},
  journal={1},
  year={2023},
  url={https://api.semanticscholar.org/CorpusID:257675337}
}

@inproceedings{fernandez2023three,
  title={Three bricks to consolidate watermarks for large language models},
  author={Fernandez, Pierre and Chaffin, Antoine and Tit, Karim and Chappelier, Vivien and Furon, Teddy},
  booktitle={2023 IEEE international workshop on information forensics and security (WIFS)},
  pages={1--6},
  year={2023},
  organization={IEEE}
}

@inproceedings{sun2023codemark,
  title={Codemark: Imperceptible watermarking for code datasets against neural code completion models},
  author={Sun, Zhensu and Du, Xiaoning and Song, Fu and Li, Li},
  booktitle={Proceedings of the 31st ACM joint European software engineering conference and symposium on the foundations of software engineering},
  pages={1561--1572},
  year={2023}
}

@article{gu2023learnability,
  title={On the learnability of watermarks for language models},
  author={Gu, Chenchen and Li, Xiang Lisa and Liang, Percy and Hashimoto, Tatsunori},
  journal={arXiv preprint arXiv:2312.04469},
  year={2023}
}

@article{kuditipudi2023robust,
  title={Robust distortion-free watermarks for language models},
  author={Kuditipudi, Rohith and Thickstun, John and Hashimoto, Tatsunori and Liang, Percy},
  journal={arXiv preprint arXiv:2307.15593},
  year={2023}
}

@article{hu2023unbiased,
  title={Unbiased watermark for large language models},
  author={Hu, Zhengmian and Chen, Lichang and Wu, Xidong and Wu, Yihan and Zhang, Hongyang and Huang, Heng},
  journal={arXiv preprint arXiv:2310.10669},
  year={2023}
}

@article{Zhao2023ProvableRW,
  title={Provable Robust Watermarking for AI-Generated Text},
  author={Xuandong Zhao and Prabhanjan Vijendra Ananth and Lei Li and Yu-Xiang Wang},
  journal={ArXiv},
  year={2023},
  volume={abs/2306.17439},
  url={https://api.semanticscholar.org/CorpusID:259308864}
}

@inproceedings{christ2024undetectable,
  title={Undetectable watermarks for language models},
  author={Christ, Miranda and Gunn, Sam and Zamir, Or},
  booktitle={The Thirty Seventh Annual Conference on Learning Theory},
  pages={1125--1139},
  year={2024},
  organization={PMLR}
}

@misc{brookingswatermark,
      title={Detecting AI fingerprints: A guide to watermarking and beyond}, 
      author={Siddarth Srinivasan},
      year={2024},
      url={https://www.brookings.edu/articles/detecting-ai-fingerprints-a-guide-to-watermarking-and-beyond/}, 
}

@article{Suvra2023ASO,
  title={A Survey on the Possibilities \& Impossibilities of AI-generated Text Detection},
  author={Soumya Suvra and Ghosal and Souradip Chakraborty and Jonas Geiping and Furong Huang and Dinesh Manocha and A. S. Bedi},
  journal={Trans. Mach. Learn. Res.},
  year={2023},
  volume={2023},
  url={https://api.semanticscholar.org/CorpusID:266982110}
}

@misc{liu2024semanticinvariantrobustwatermark,
      title={A Semantic Invariant Robust Watermark for Large Language Models}, 
      author={Aiwei Liu and Leyi Pan and Xuming Hu and Shiao Meng and Lijie Wen},
      year={2024},
      eprint={2310.06356},
      archivePrefix={ICLR},
      primaryClass={cs.CR},
      url={https://arxiv.org/abs/2310.06356}, 
}

@misc{gu2025invisibleentropysafeefficient,
      title={Invisible Entropy: Towards Safe and Efficient Low-Entropy LLM Watermarking}, 
      author={Tianle Gu and Zongqi Wang and Kexin Huang and Yuanqi Yao and Xiangliang Zhang and Yujiu Yang and Xiuying Chen},
      year={2025},
      eprint={2505.14112},
      archivePrefix={arXiv},
      primaryClass={cs.CL},
      url={https://arxiv.org/abs/2505.14112}, 
}

@inproceedings{papineni-etal-2002-bleu,
    title = "{B}leu: a Method for Automatic Evaluation of Machine Translation",
    author = "Papineni, Kishore  and
      Roukos, Salim  and
      Ward, Todd  and
      Zhu, Wei-Jing",
    editor = "Isabelle, Pierre  and
      Charniak, Eugene  and
      Lin, Dekang",
    booktitle = "Proceedings of the 40th Annual Meeting of the Association for Computational Linguistics",
    month = jul,
    year = "2002",
    address = "Philadelphia, Pennsylvania, USA",
    publisher = "Association for Computational Linguistics",
    url = "https://aclanthology.org/P02-1040/",
    doi = "10.3115/1073083.1073135",
    pages = "311--318"
}

@inproceedings{lin-2004-rouge,
    title = "{ROUGE}: A Package for Automatic Evaluation of Summaries",
    author = "Lin, Chin-Yew",
    booktitle = "Text Summarization Branches Out",
    month = jul,
    year = "2004",
    address = "Barcelona, Spain",
    publisher = "Association for Computational Linguistics",
    url = "https://aclanthology.org/W04-1013/",
    pages = "74--81"
}

@inproceedings{rei-etal-2020-comet,
    title = "{COMET}: A Neural Framework for {MT} Evaluation",
    author = "Rei, Ricardo  and
      Stewart, Craig  and
      Farinha, Ana C  and
      Lavie, Alon",
    editor = "Webber, Bonnie  and
      Cohn, Trevor  and
      He, Yulan  and
      Liu, Yang",
    booktitle = "Proceedings of the 2020 Conference on Empirical Methods in Natural Language Processing (EMNLP)",
    month = nov,
    year = "2020",
    address = "Online",
    publisher = "Association for Computational Linguistics",
    url = "https://aclanthology.org/2020.emnlp-main.213/",
    doi = "10.18653/v1/2020.emnlp-main.213",
    pages = "2685--2702"
}

@inproceedings{reimers-2019-sentence-bert,
    title = "Sentence-BERT: Sentence Embeddings using Siamese BERT-Networks",
    author = "Reimers, Nils and Gurevych, Iryna",
    booktitle = "Proceedings of the 2019 Conference on Empirical Methods in Natural Language Processing",
    month = "11",
    year = "2019",
    publisher = "Association for Computational Linguistics",
    url = "https://arxiv.org/abs/1908.10084",
}

@misc{touvron2023llama2openfoundation,
      title={Llama 2: Open Foundation and Fine-Tuned Chat Models}, 
      author={Hugo Touvron and Louis Martin and Kevin Stone and Peter Albert and Amjad Almahairi and Yasmine Babaei and Nikolay Bashlykov and Soumya Batra and Prajjwal Bhargava and Shruti Bhosale and Dan Bikel and Lukas Blecher and Cristian Canton Ferrer and Moya Chen and Guillem Cucurull and David Esiobu and Jude Fernandes and Jeremy Fu and Wenyin Fu and Brian Fuller and Cynthia Gao and Vedanuj Goswami and Naman Goyal and Anthony Hartshorn and Saghar Hosseini and Rui Hou and Hakan Inan and Marcin Kardas and Viktor Kerkez and Madian Khabsa and Isabel Kloumann and Artem Korenev and Punit Singh Koura and Marie-Anne Lachaux and Thibaut Lavril and Jenya Lee and Diana Liskovich and Yinghai Lu and Yuning Mao and Xavier Martinet and Todor Mihaylov and Pushkar Mishra and Igor Molybog and Yixin Nie and Andrew Poulton and Jeremy Reizenstein and Rashi Rungta and Kalyan Saladi and Alan Schelten and Ruan Silva and Eric Michael Smith and Ranjan Subramanian and Xiaoqing Ellen Tan and Binh Tang and Ross Taylor and Adina Williams and Jian Xiang Kuan and Puxin Xu and Zheng Yan and Iliyan Zarov and Yuchen Zhang and Angela Fan and Melanie Kambadur and Sharan Narang and Aurelien Rodriguez and Robert Stojnic and Sergey Edunov and Thomas Scialom},
      year={2023},
      eprint={2307.09288},
      archivePrefix={arXiv},
      primaryClass={cs.CL},
      url={https://arxiv.org/abs/2307.09288}, 
}

@misc{gemma2_2024,
      title={{Gemma 2: Improving Open Language Models at a Practical Size}}, 
      author={{Gemma Team}},
      year={2024},
      eprint={2408.00118},
      archivePrefix={arXiv},
      primaryClass={cs.CL},
      url={https://arxiv.org/abs/2408.00118},
}

@inproceedings{feng-etal-2022-language,
    title = "Language-agnostic {BERT} Sentence Embedding",
    author = "Feng, Fangxiaoyu  and
      Yang, Yinfei  and
      Cer, Daniel  and
      Arivazhagan, Naveen  and
      Wang, Wei",
    editor = "Muresan, Smaranda  and
      Nakov, Preslav  and
      Villavicencio, Aline",
    booktitle = "Proceedings of the 60th Annual Meeting of the Association for Computational Linguistics (Volume 1: Long Papers)",
    month = may,
    year = "2022",
    address = "Dublin, Ireland",
    publisher = "Association for Computational Linguistics",
    url = "https://aclanthology.org/2022.acl-long.62/",
    doi = "10.18653/v1/2022.acl-long.62",
    pages = "878--891"
}

@misc{alma7b,
      title={A Paradigm Shift in Machine Translation: Boosting Translation Performance of Large Language Models}, 
      author={Haoran Xu and Young Jin Kim and Amr Sharaf and Hany Hassan Awadalla},
      year={2023},
      eprint={2309.11674},
      archivePrefix={arXiv},
      primaryClass={cs.CL}
}

@article{c4realnews,
  author  = {Colin Raffel and Noam Shazeer and Adam Roberts and Katherine Lee and Sharan Narang and Michael Matena and Yanqi Zhou and Wei Li and Peter J. Liu},
  title   = {Exploring the Limits of Transfer Learning with a Unified Text-to-Text Transformer},
  journal = {Journal of Machine Learning Research},
  year    = {2020},
  volume  = {21},
  number  = {140},
  pages   = {1--67},
  url     = {http://jmlr.org/papers/v21/20-074.html}
}

@inproceedings{narayan-etal-2018-dont,
    title = "Don{'}t Give Me the Details, Just the Summary! Topic-Aware Convolutional Neural Networks for Extreme Summarization",
    author = "Narayan, Shashi  and
      Cohen, Shay B.  and
      Lapata, Mirella",
    editor = "Riloff, Ellen  and
      Chiang, David  and
      Hockenmaier, Julia  and
      Tsujii, Jun{'}ichi",
    booktitle = "Proceedings of the 2018 Conference on Empirical Methods in Natural Language Processing",
    month = oct # "-" # nov,
    year = "2018",
    address = "Brussels, Belgium",
    publisher = "Association for Computational Linguistics",
    url = "https://aclanthology.org/D18-1206/",
    doi = "10.18653/v1/D18-1206",
    pages = "1797--1807"
}

@inproceedings{kocmi-etal-2023-findings,
    title = "Findings of the 2023 Conference on Machine Translation ({WMT}23): {LLM}s Are Here but Not Quite There Yet",
    author = "Kocmi, Tom  and
      Avramidis, Eleftherios  and
      Bawden, Rachel  and
      Bojar, Ond{\v{r}}ej  and
      Dvorkovich, Anton  and
      Federmann, Christian  and
      Fishel, Mark  and
      Freitag, Markus  and
      Gowda, Thamme  and
      Grundkiewicz, Roman  and
      Haddow, Barry  and
      Koehn, Philipp  and
      Marie, Benjamin  and
      Monz, Christof  and
      Morishita, Makoto  and
      Murray, Kenton  and
      Nagata, Masaaki  and
      Nakazawa, Toshiaki  and
      Popel, Martin  and
      Popovi{\'c}, Maja  and
      Shmatova, Mariya  and
      Suzuki, Jun",
    editor = "Koehn, Philipp  and
      Haddow, Barry  and
      Kocmi, Tom  and
      Monz, Christof",
    booktitle = "Proceedings of the Eighth Conference on Machine Translation",
    month = dec,
    year = "2023",
    address = "Singapore",
    publisher = "Association for Computational Linguistics",
    url = "https://aclanthology.org/2023.wmt-1.1/",
    doi = "10.18653/v1/2023.wmt-1.1",
    pages = "1--42"
}

@inproceedings{wmt19translate,
  author = {Bojar, Ondřej and Graham, Yvette and Monz, Christof and Post, Matt and Bojar, Ondřej and Graham, Yvette and Monz, Christof and Post, Matt},
  title = {ACL 2019 Fourth Conference on Machine Translation (WMT19), Shared Task: Machine Translation of News},
  year = {2019},
  url = {http://www.statmt.org/wmt19/translation-task.html}
}

@inproceedings{indykandmotwani1998,
author = {Indyk, Piotr and Motwani, Rajeev},
title = {Approximate nearest neighbors: towards removing the curse of dimensionality},
year = {1998},
isbn = {0897919629},
publisher = {Association for Computing Machinery},
address = {New York, NY, USA},
url = {https://doi.org/10.1145/276698.276876},
doi = {10.1145/276698.276876},
booktitle = {Proceedings of the Thirtieth Annual ACM Symposium on Theory of Computing},
pages = {604–613},
numpages = {10},
location = {Dallas, Texas, USA},
series = {STOC '98}
}

@inproceedings{charikar2002similarity,
  title={{S}imilarity {E}stimation {T}echniques from {R}ounding {A}lgorithms},
  author={Charikar, Moses},
  booktitle={Proceedings of the thirty-fourth annual ACM symposium on Theory of computing},
  pages={380--388},
  year={2002},
  organization={ACM}
}

@inproceedings{lee-etal-2024-wrote,
    title = "Who Wrote this Code? Watermarking for Code Generation",
    author = "Lee, Taehyun  and
      Hong, Seokhee  and
      Ahn, Jaewoo  and
      Hong, Ilgee  and
      Lee, Hwaran  and
      Yun, Sangdoo  and
      Shin, Jamin  and
      Kim, Gunhee",
    editor = "Ku, Lun-Wei  and
      Martins, Andre  and
      Srikumar, Vivek",
    booktitle = "Proceedings of the 62nd Annual Meeting of the Association for Computational Linguistics (Volume 1: Long Papers)",
    month = aug,
    year = "2024",
    address = "Bangkok, Thailand",
    publisher = "Association for Computational Linguistics",
    url = "https://aclanthology.org/2024.acl-long.268/",
    doi = "10.18653/v1/2024.acl-long.268",
    pages = "4890--4911"
}

@InProceedings{pmlr-v202-kirchenbauer23a,
  title = 	 {A Watermark for Large Language Models},
  author =       {Kirchenbauer, John and Geiping, Jonas and Wen, Yuxin and Katz, Jonathan and Miers, Ian and Goldstein, Tom},
  booktitle = 	 {Proceedings of the 40th International Conference on Machine Learning},
  pages = 	 {17061--17084},
  year = 	 {2023},
  editor = 	 {Krause, Andreas and Brunskill, Emma and Cho, Kyunghyun and Engelhardt, Barbara and Sabato, Sivan and Scarlett, Jonathan},
  volume = 	 {202},
  series = 	 {Proceedings of Machine Learning Research},
  month = 	 {23--29 Jul},
  publisher =    {PMLR},
  url = 	 {https://proceedings.mlr.press/v202/kirchenbauer23a.html}
}

@inproceedings{hou-etal-2024-semstamp,
    title = "{S}em{S}tamp: A Semantic Watermark with Paraphrastic Robustness for Text Generation",
    author = "Hou, Abe  and
      Zhang, Jingyu  and
      He, Tianxing  and
      Wang, Yichen  and
      Chuang, Yung-Sung  and
      Wang, Hongwei  and
      Shen, Lingfeng  and
      Van Durme, Benjamin  and
      Khashabi, Daniel  and
      Tsvetkov, Yulia",
    editor = "Duh, Kevin  and
      Gomez, Helena  and
      Bethard, Steven",
    booktitle = "Proceedings of the 2024 Conference of the North American Chapter of the Association for Computational Linguistics: Human Language Technologies (Volume 1: Long Papers)",
    month = jun,
    year = "2024",
    address = "Mexico City, Mexico",
    publisher = "Association for Computational Linguistics",
    url = "https://aclanthology.org/2024.naacl-long.226/",
    doi = "10.18653/v1/2024.naacl-long.226",
    pages = "4067--4082"
}

@inproceedings{lu-etal-2024-entropy,
    title = "An Entropy-based Text Watermarking Detection Method",
    author = "Lu, Yijian  and
      Liu, Aiwei  and
      Yu, Dianzhi  and
      Li, Jingjing  and
      King, Irwin",
    editor = "Ku, Lun-Wei  and
      Martins, Andre  and
      Srikumar, Vivek",
    booktitle = "Proceedings of the 62nd Annual Meeting of the Association for Computational Linguistics (Volume 1: Long Papers)",
    month = aug,
    year = "2024",
    address = "Bangkok, Thailand",
    publisher = "Association for Computational Linguistics",
    url = "https://aclanthology.org/2024.acl-long.630/",
    doi = "10.18653/v1/2024.acl-long.630",
    pages = "11724--11735"
}

@misc{takezawa2025necessarysufficientwatermarklarge,
      title={Necessary and Sufficient Watermark for Large Language Models}, 
      author={Yuki Takezawa and Ryoma Sato and Han Bao and Kenta Niwa and Makoto Yamada},
      year={2025},
      eprint={2310.00833},
      archivePrefix={arXiv},
      primaryClass={cs.CL},
      url={https://arxiv.org/abs/2310.00833}, 
}

@inproceedings{hou-etal-2024-k,
    title = "k-{S}em{S}tamp: A Clustering-Based Semantic Watermark for Detection of Machine-Generated Text",
    author = "Hou, Abe  and
      Zhang, Jingyu  and
      Wang, Yichen  and
      Khashabi, Daniel  and
      He, Tianxing",
    editor = "Ku, Lun-Wei  and
      Martins, Andre  and
      Srikumar, Vivek",
    booktitle = "Findings of the Association for Computational Linguistics: ACL 2024",
    month = aug,
    year = "2024",
    address = "Bangkok, Thailand",
    publisher = "Association for Computational Linguistics",
    url = "https://aclanthology.org/2024.findings-acl.98/",
    doi = "10.18653/v1/2024.findings-acl.98",
    pages = "1706--1715"
}

@inproceedings{ajith-etal-2024-downstream,
    title = "Downstream Trade-offs of a Family of Text Watermarks",
    author = "Ajith, Anirudh  and
      Singh, Sameer  and
      Pruthi, Danish",
    editor = "Al-Onaizan, Yaser  and
      Bansal, Mohit  and
      Chen, Yun-Nung",
    booktitle = "Findings of the Association for Computational Linguistics: EMNLP 2024",
    month = nov,
    year = "2024",
    address = "Miami, Florida, USA",
    publisher = "Association for Computational Linguistics",
    url = "https://aclanthology.org/2024.findings-emnlp.821/",
    doi = "10.18653/v1/2024.findings-emnlp.821",
    pages = "14039--14053"
}

\appendix
\section{Appendix}

\subsection{Hyperparameter Sweep for Translation and Summarization} 
\label{app:hyperparams}

Given different sets of hyperparameters for each watermark methods, to ensure apple-to-apple comparisons we sweep over a range of values of each method's relevant hyperparameters (Table \ref{tab:hyperparams}). For each task, we sample 100 evaluation samples and run these watermark hyperparameter configs on these subset to obtain the Pareto curves in Figure \ref{fig:pareto}. The best configs are then used to run the full evaluation set reported in Tables \ref{tab:constrained}, \ref{tab:code_and_c4}, and \ref{tab:wmt23}.

\begin{table*}[!h]
    \centering
    \small
    \begin{tabular}{c|c|c}
    \toprule
    \textbf{Hyperparmeter} & \textbf{Methods} &  \textbf{Values}\\
    \midrule
    Text generation temperature $t$ & \kgw, \semstamp, \ksemstamp, \ours & [0.7, 0.85, 1.0, 1.2, 1.5] \\
    Green/red list and regions ratio $\gamma$ & \kgw, \semstamp, \ksemstamp, \ours & [0.1, 0.25, 0.5, 0.75] \\
    Logit bias $\delta$ & \kgw & [0.1, 0.5, 1.0, 2.0, 4.0] \\
    LSH dimension $n$ & \semstamp, \ksemstamp, \ours & [2, 3, 4, 5] \\
    High-quality cluster samples $c$ & \ours & [50] \\
    \bottomrule
    \end{tabular}
    \caption{Relevant hyperparameter for watermark methods}
    \label{tab:hyperparams}
\end{table*}

\subsection{Full Experimental Results} 
\label{app:full}

As mentioned in \S \ref{sec:main_experiment}, for sentence translation, summarization, and code generation, we reported precision, recall, \fscore{} and not the usual AUROC / TP@FPR=1 / TP@FPR=5. This is due to the nature of the evaluated tasks: since we only generate a handful (often single) sequences per example, the z-test employed in sequence-level watermarking SemStamp is severely underpowered. In fact, sequence-level watermark detection for these tasks is effectively binary classification: the sentence is watermarked or not. Therefore, the standard AUROC metric, where the area is computed by sweeping over different classification thresholds, is ill-suited for tasks that only contain a single classification threshold. This is not a problem for token-level watermarking, however, since there are enough tokens for the z-test. To ensure fairness of comparison between the token-level baselines and sentence-level approaches, we select the threshold that yields the highest true positive rate / false positive rate ratio and reported the corresponding precision, recall, and \fscore. 

\begin{figure*}[h]
    \centering
    \begin{subfigure}{0.32\textwidth}
        \centering
        \includegraphics[width=\linewidth]{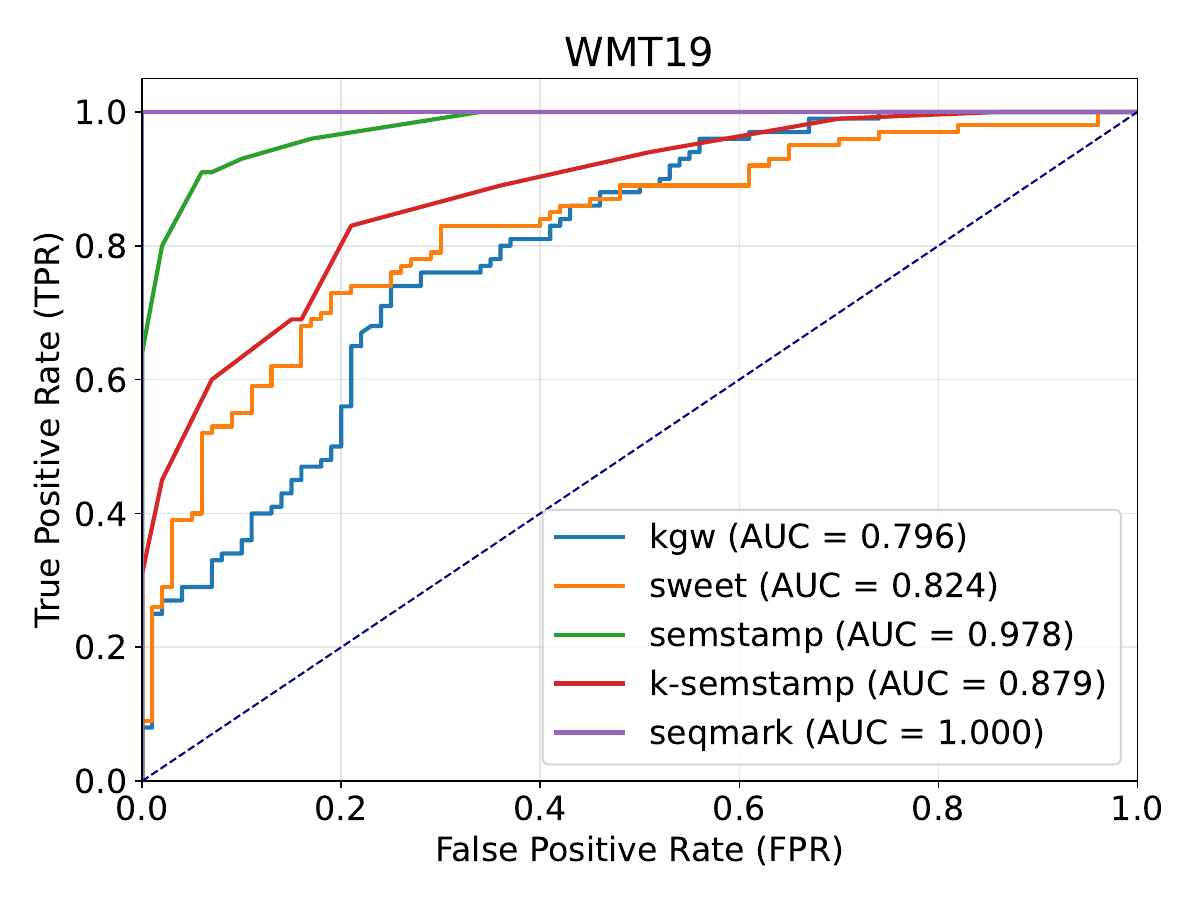} 
    \end{subfigure}
    \begin{subfigure}{0.32\textwidth}
        \centering
        \includegraphics[width=\linewidth]{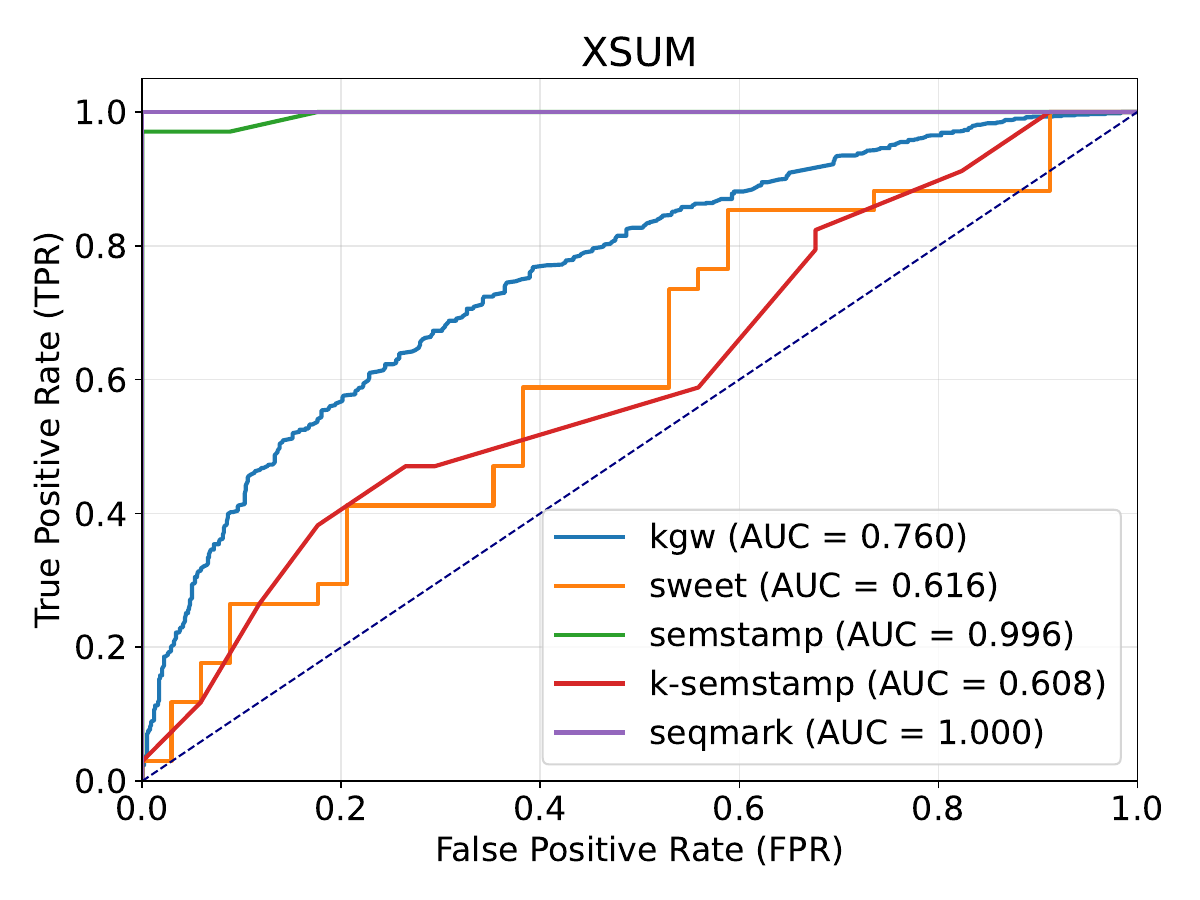}
    \end{subfigure}
    \begin{subfigure}{0.32\textwidth}
        \centering
        \includegraphics[width=\linewidth]{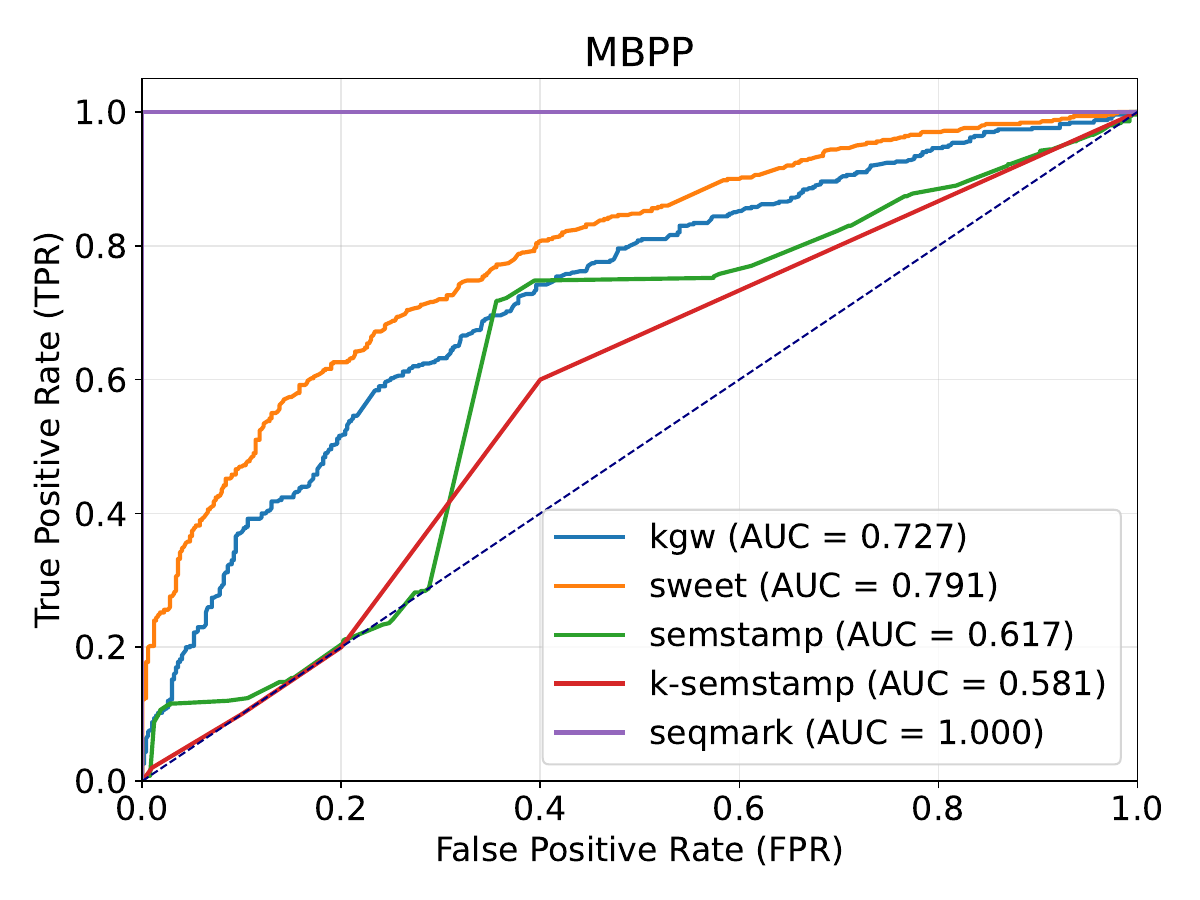}
    \end{subfigure}
    \caption{Aggregated ROC across different classification thresholds for sentence translation (right), summarization (middle), and code generation (left).}. 
    \vspace{-5mm}
    \label{fig:roc}
\end{figure*}

To ensure the robustness of our study, we also experiment with \textit{aggregated AUROC}, where instead of considering per-example sequence, we aggregate the sequences across multiple examples of the evaluation set. Specifically, each new example now consists of 30 sequences across consecutive examples. For example, in translation, if there are 3000 examples where each example generates one sentence translation, then we aggregate across 30 examples, for a total of 100 data points. We then compute the AUROC / TP@1 / TP@5 for this new evaluation set. Results in Table \ref{tab:translation_full} and \ref{tab:mbpp_full} show consistency between aggregated AUROC and \fscore{}. In fact, our approach SeqMark achieves perfect classification score under aggregated AUROC. We also plot the aggregated ROC across different thresholds in Figure \ref{fig:roc}.

\begin{table*}[!h]
\centering
\small
\setlength{\tabcolsep}{3pt}
\renewcommand{\arraystretch}{1.15}
\resizebox{\textwidth}{!}{%
\begin{tabular}{lcccccc}
\toprule
& \multicolumn{3}{c}{\textbf{Translation (WMT19 De-En)}} 
& \multicolumn{3}{c}{\textbf{Summarization (XSum)}} \\
\cmidrule(lr){2-4} \cmidrule(lr){5-7} 
& \textbf{COMET} $\uparrow$
& \textbf{P / R / \fscore (h)}
& \textbf{agg-AUROC / TP@1 / TP@5}
& \textbf{R-L} $\uparrow$ / \textbf{COMET} $\uparrow$
& \textbf{P / R / \fscore (h)}
& \textbf{agg-AUROC / TP@1 / TP@5} \\
\midrule
\textit{No Watermark} 
& \textit{87.4} 
& - 
& - 
& \textit{20.5 / 69.0}
& - 
& - \\
\kgw 
& 87.4 
& 57.2 / 36.6 / 45.7 
& 79.6 / 25.0 / 29.0
& 20.1 / 68.8
& 85.2 / 30.1 / 44.9
& 75.2 / 26.4 / 45.8 \\
\sweet
& 87.2
& 57.5 / 30.0 / 39.5
& 82.4 / 20.0 / 47.0
& 18.5 / 68.7
& 52.4 / 49.9 / 51.1
& 52.8 / 2.9 / 10.7 \\
\texttt{STA-1}
& 87.1
& 60.6 / 30.1 / 40.2	
& 59.0 / 5.2 / 14.0
& -
& -
& - \\
\semstamp
& 87.4
& 59.5 / 73.7 / 65.9
& 97.8 / 63.0 / 91.0
& 20.0 / 68.7
& 70.6 / 52.8 / 60.4
& 99.6 / 76.4 / 91.2 \\
\ksemstamp
& \textbf{87.5}
& 63.3 / 33.0 / 43.4
& 87.9 / 48.0 / 55.0
& 20.7 / \textbf{68.9}
& 54.2 / 22.0 / 31.3
& 60.8 / 3.9 / 11.8 \\
\ours
& 87.1
& \textbf{76.9 / 77.3 / 77.1}
& \textbf{100 / 100 / 100}
& \textbf{21.6} / 68.5
& \textbf{81.3 / 100 / 89.7}
& \textbf{100 / 100 / 100} \\
\bottomrule
\end{tabular}
}
\caption{Results on precision, recall, \fscore{} and aggregated AUROC metrics for sentence translation and summarization. We evaluated with human generations as negative examples. Best results are \textbf{bold}. We also include an additional baseline STA-1 for translation \citep{mao-etal-2025-watermarking}}

\label{tab:translation_full}
\end{table*}

\begin{table*}[!h]
\centering
\small 
\setlength{\tabcolsep}{3pt} 
\renewcommand{\arraystretch}{1.15} 
\resizebox{0.5\textwidth}{!}{%
\begin{tabular}{lccc}
\toprule
\textbf{Approach} & \textbf{pass@1} & \textbf{P / R / \fscore (h)} & \textbf{agg-AUROC / TP@1 / TP@5}\\
\midrule
\textit{No Watermark} & \textit{33.8} & - & - \\
\kgw        & 15.0 & 87.5 / 31.0 / 45.7 & 71.2 / 8.0 / 17.0\\
\sweet & 33.2 & 79.5 / 62.3 / 70.2 & 80.5 / 63.2 / 71.6\\
\texttt{STA-1} & 34.0 & 55.4 / 21.6 / 31.1 & 54.5 / 8.6 / 17.4\\
\semstamp    &  \textbf{34.2} & 66.9 / 71.8 / 69.2 & 61.7 / 8.0 / 12.0\\
\ksemstamp & 34.0 & 50.0 / 98.7 / 66.5 & 58.1 / 2.0 / 5.0\\
\ours & 33.6 & \textbf{76.0 / 86.0 / 80.7} & \textbf{100 / 100 / 100}\\
\bottomrule
\end{tabular}
}
\caption{Results on precision, recall, \fscore{} and aggregated AUROC metrics for code generation. We also include an additional baseline for low-entropy texts STA-1 \citep{mao-etal-2025-watermarking}.}
\vspace{-2mm}
\label{tab:mbpp_full}
\end{table*}


\subsection{Detecting Watermarked LLM against other LMs} 
In addition to testing the ability of watermarking schemes to differentiate between watermarked texts and texts generated either by humans or non-watermarked LM, we also report results on the ability to differentiate between watermarked texts and texts generated from other (non-watermarked) language models.
Using 100 samples from WMT-19 German-English as test bed, we compare different watermarking methods on ALMA-7B with unwatermarked completions from ALMA-7B, Gemma-2-4B instruct, and Llama-2-7B-Chat. Results from Table \ref{tab:lm_evaluation} indicate that SeqMark differentiates the watermarked text of a particular LM from other text sequences, be it human-generated or other LM-generated equally well. 

\subsection{Choice of Sentence Encoders} 
\label{app:encoder}

\begin{table*}[!h]
\centering
\small
\setlength{\tabcolsep}{2pt}
\renewcommand{\arraystretch}{1.2}
\begin{tabular}{lcc}
\toprule
\textbf{Method (Encoder)} & \textbf{ROUGE-L / COMET} & \textbf{P / R / F1 (h)}\\
\hline
\semstamp (all-mpnet-base-v1) & 21.1 / 68.7 & 70.6 / 52.8 / 60.4 \\
\ours (all-mpnet-base-v1)  & 20.6 / 68.5 & 76.1 / 100 / 86.4  \\
\semstamp (cde-small-v2)      & 22.0 / 68.3 & 80.0 / 48.8 / 60.6 \\
\ours (cde-small-v2)       & 21.2 / 68.3 & 79.5 / 100 / 88.6  \\
\bottomrule
\end{tabular}
\caption{Comparison of methods across different encoders.}
\label{tab:encoders}
\end{table*}

Our choice of sentence encoder was task-dependent: LABSE encoder for machine translation and SBERT for summarization. We also experimented with other encoders \citep{morris2024contextualdocumentembeddings} and did not see major variations in results, indicating the robustness of SeqMark across different sentence encoders (Table \ref{tab:encoders}).

\begin{table}[!t]
\centering
\small 
\setlength{\tabcolsep}{3pt} 
\renewcommand{\arraystretch}{1.15} 
\begin{tabular}{lccccc}
\toprule
\tiny \textbf{Approach} & \tiny \textbf{COMET $\uparrow$} & \tiny \textbf{Human}  & \tiny \textbf{ALMA}  & \tiny \textbf{Gemma-2}  & \tiny \textbf{Llama-2}\\
\midrule
\kgw        & \textbf{85.5} & 47.8 & 47.6 & 42.8 & 43.2 \\
\semstamp    & 85.2 & 65.5 & 65.5 & 66.7 & 67.6\\
\ours & 85.0 & \textbf{80.2} & \textbf{79.0} & \textbf{78.7} & \textbf{81.8} \\
\bottomrule
\end{tabular}
\caption{\fscore{} on WMT19 translation task testing detection against negative examples from various sources: human and other non-watermarked LLMs.}
\label{tab:lm_evaluation}
\end{table}

\subsection{Discussion on Paraphrasing Attacks}
\label{app:paraphrase}

A potential limitation of our approach is that, unlike SemStamp / k-SemStamp, it might be susceptible to paraphrasing attacks. By design, SeqMark spreads similar generations more evenly in the semantic space; thus a strong paraphraser can potentially evade the watermark by transforming the generation into another semantic region. However, in our preliminary paraphrase attack experiments following SemStamp, we observe that the paraphrasing outputs often have low text-quality, indicating that current paraphrasing models are not strong enough to maintain text quality, especially for low-entropy tasks like translations (Table \ref{tab:paraphrase}). In addition, existing sentence-level algorithms e.g. SemStamp does not have strong watermarking performance on these tasks, even without paraphrasing attacks (Table \ref{tab:constrained}). 

\begin{table*}[t]
\centering
\small
\begin{tabular}{lcc}
\toprule
\textbf{Method}  & \textbf{COMET} \\
\midrule
\kgw & 85.5 \\
\kgw (paraphrase attacked)  & 79.3 \\
\midrule
\semstamp & 85.2 \\
\semstamp (paraphrase attacked) & 75.4 \\
\midrule
\ours & 85.0 \\
\ours (paraphrase attacked)  & 80.1 \\
\bottomrule
\end{tabular}
\caption{Generation quality under paraphrasing attacks, for sentence translation task.}
\label{tab:paraphrase}
\end{table*}

\subsection{Imperceptibility when Watermarking Translation Outputs}
\label{app:imperceptible}

\citet{takezawa2025necessarysufficientwatermarklarge} recently showed that watermarking with minimal intervention is possible for machine translation tasks via their token-level NS-Watermarking method. When compared to our approach, we observed that this method achieves near perfect watermark detection performance but suffers in terms of quality measured by COMET with $\sim 6\%$ absolute decrease. Additionally, they acknowledge that by design their approach is not imperceptible -- it is easy to tell if the output has been watermarked. Looking at Table~\ref{tab:ns_watermark_qualitative}, it is easy to notice the minimal yet awkward token choices for watermarked translation. Concerningly, this imperceptibility makes this approach susceptible to simple post-editing attacks. On the other hand, our approach results in imperceptible watermarked generation, making it robust to simple attacks. 

\subsection{Proof of Theorem 1}
\label{app:performance_proof}

In this section, we aim to show that SeqMark mean-centering transformation improves detection accuracy for LSH-based sentence-level watermarking, assuming the sentence embeddings are similar to each other (Theorem 1).

\paragraph{Preliminaries} We first define the relevant preliminaries for our analysis

\begin{itemize}
    \item \textbf{Sequence Encoding}: Given $n$ candidate sequences $c_1,..., c_n$, sentence embedding $v_i$ is produced using sentence encoder $E$: $\mathbf{x_i} = E(c_i) \in \mathbb{R}^h, \forall 1 \leq i \leq n$
    \item \textbf{Locality Sensitive Hasing for Text Watermarking} \citep{hou-etal-2024-semstamp}: Given dimension $h$, we sample $d$ random vectors from a multivariate normal distribution (i.e. $\mathbf{g}^{(i)} \sim \mathcal{N}(\mathbf{0}, I_h), \forall 0 \leq i \leq d$) to specify $d$ random hyperplanes and thus $2^d$ semantic regions. For sentence-level watermarking, we are given a "valid region ratio" $\gamma$ denoting the fraction of valid regions.
    \item \textbf{LSH Collision Probability} \citep{charikar2002similarity}: given two vectors $\mathbf{x_i},\mathbf{x_j} \in \mathbb{R}^h$ with angle $\theta_{ij} \in [0, \pi]$, the probability that they lie in the same region (i.e. having the same LSH signature) is: $$P(\sign(\mathbf{x_i}) = \sign(\mathbf{y_i})) = \left( 1-\frac{\theta_{ij}}{\pi}\right)^d$$
    where $\sign(\mathbf{x})$ is the concatenation of the dot product between $\mathbf{x}$ and all the normal vectors $\mathbf{g}^{(i)}$. For simplicity, let $q_{ij} = \left( 1-\frac{\theta_{ij}}{\pi}\right)^d$
    \item \textbf{Detection Accuracy} An important quantity for our analysis is the indicator $I^{(x)}$, where $I^{(x)}=1$ \textit{if there exists at least one vector $\mathbf{x_i}$ in the valid regions}, and 0 otherwise. This quantity is the watermark detection accuracy of a single example, since $I^{(x)}=1$ if there exists at least one sentence embedding $\mathbf{x_i}$ that lies in the valid region.
\end{itemize}

\paragraph{Proposition 1} Given $n$ vectors $\mathbf{x_1},...,\mathbf{x_n} \in \mathbb{R}^h$, $2^d$ semantic regions partitioned using LSH, and $\gamma$ ratio of valid regions, we can approximate the average detection accuracy $E[I]$ as:

$$E[I^{(x)}] \approx n\gamma - \binom{n}{2}\gamma^2 - \gamma(1-\gamma)\sum_{i<j}\left( 1-\frac{\theta_{ij}}{\pi}\right)^d$$

\paragraph{Proof} Let $A_i$ be the event that vector $x_i$ lands in a valid region. We then have: 
\[
\begin{split}
        E[I^{(x)}] &= P(\cup_i A_i) \\ 
        &= \sum_i P(A_i) - \sum_{i<j} P(A_i \cap A_j) \\
        &= + \sum_{i<j<k} P(A_i \cap A_j \cap A_k) - ... \\
        &\quad \text{(inclusion-exclusion principle)}\\ 
\end{split}
\]
where $P(A_i)$ denotes the probability that vector $\mathbf{x_i}$ lands in a valid region, $P(A_i \cap A_j)$ denotes the probability that vectors $\mathbf{x_i}$, $\mathbf{x_j}$ land in valid regions (which can be the same or different), etc.

A reasonable assumption is that the higher-order terms are smaller in mass (give a one line reason). To simplify the computation, we make the second-order pairwise approximation (i.e. keeping the only two terms):
$$E[I^{(x)}] \approx \sum_i P(A_i) - \sum_{i<j} P(A_i \cap A_j)$$

Now we compute $P(A_i)$ and $P(A_i \cap A_j)$. The probability that a vector $\mathbf{x_i}$ lands in a valid region is $P(A_i) = \gamma$. For $P(A_i \cap A_j)$, we have:
\[
\begin{aligned}
P(A_i \cap A_j)
&= P(A_i \cap A_j \mid s_i = s_j) P(s_i = s_j) \\
&\quad + P(A_i \cap A_j \mid s_i \neq s_j) P(s_i \neq s_j) \\
&= \gamma q_{ij} + \gamma^2 (1 - q_{ij}) \\
&= \gamma^2 + \gamma(1 - \gamma) q_{ij},
\end{aligned}
\]
where $s_k \equiv \sign(\mathbf{x}_k)$.
Putting everything together, we have 
\[
\begin{split}
E[I^{(x)}] &\approx n\gamma - \binom{n}{2}\gamma^2 - \gamma(1-\gamma)\sum_{i<j}q_{ij} \\ 
&\approx n\gamma - \binom{n}{2}\gamma^2 - \gamma(1-\gamma)\sum_{i<j}\left( 1-\frac{\theta_{ij}}{\pi}\right)^d
\end{split}
\]

\hfill $\square$

Proposition 1 nicely relates $E[I^{(x)}]$ to the pairwise angles between vectors $\theta_{ij}$. From here we can reason for various value of $\theta_{ij}$. 

\paragraph{Collorary 1} $E[I^{(x)}]$ monotonically nondecreases with respect to $\theta_{ij}$, for all $1 \leq i,j \leq n$

\paragraph{Proof} The proof basically works by showing that $dE[I^{(x)}]/d\theta_{ij} \geq 0$. Fix any pair $i,j$. The partial derivative of $E[I^{(x)}]$ with respect to 
$\theta_{ij}$ is
\[
\begin{split}
\frac{\partial E[I^{(x)}]}{\partial \theta_{ij}}
&= -\gamma(1-\gamma)\,\frac{\partial q_{ij}}{\partial \theta_{ij}} \\
&= \gamma(1-\gamma)\frac{d}{\pi}
\left(1-\frac{\theta_{ij}}{\pi}\right)^{d-1}
\end{split}
\]

All factors on the right-hand side are nonnegative for 
$\theta_{ij}\in[0,\pi]$. Therefore,
\[
\frac{\partial E[I^{(x)}]}{\partial \theta_{ij}} \ge 0,
\]
so $E[I^{(x)}]$ is monotone nondecreasing in each $\theta_{ij}$. Moreover, if 
$0<\gamma<1$, $d>0$, and $\theta_{ij}\in(0,\pi)$, then the derivative is strictly 
positive, and $E[I^{(x)}]$ is strictly increasing in $\theta_{ij}$.

\hfill $\square$

\paragraph{Theorem 1} Given $n$ vectors $\mathbf{x_1},...,\mathbf{x_n}$ such that  
$x_i = \mu + \epsilon_i$ and there exists $\delta \in [0,1]$ such that $||\epsilon_i|| \leq \delta||\mu||$. Mean-centering transformation result in vectors $\mathbf{y_i} = x_i - \frac{1}{n}\sum_i x_i = \epsilon_i$. If the following condition holds for a fixed pair $i,j$:
$$cos \theta^{(y)}_{ij} = \frac{\epsilon_i \cdot \epsilon_j}{||\epsilon_i||||\epsilon_j||} \leq \frac{1-2\delta - \delta^2}{1+2\delta+\delta^2}$$

then $E[I^{(y)}] \geq E[I^{(x)}]$

\paragraph{Proof}
By Proposition 1 and Corollary 1 it suffices to show that for every pair \((i,j)\)
\[
\cos\theta^{(y)}_{ij}\le \cos\theta^{(x)}_{ij}
\]
since \(E[I]\) is nonincreasing in \(\cos\theta_{ij}\) (equivalently nondecreasing in \(\theta_{ij}\)).

Scale $\mu$ so that $\|\mu\|=1$ (the comparison is homogeneous in $\mu$). 
Recall $x_i=\mu+\epsilon_i$ and $x_j=\mu+\epsilon_j$ with
$\|\epsilon_i\|,\|\epsilon_j\|\le\delta$
Then
\[
x_i\cdot x_j = 1 + \mu\!\cdot\!(\epsilon_i+\epsilon_j) + \epsilon_i\cdot\epsilon_j
\]
We also have
\[
\scalebox{0.9}{$
\|x_i\|\,\|x_j\|
= \sqrt{\big(1 + 2 \mu\cdot\epsilon_i + \|\epsilon_i\|^2\big)\big(1 + 2 \mu\cdot\epsilon_j + \|\epsilon_j\|^2\big)}
$}
\]

By Cauchy–Schwarz inequalities, we have the two bounds: $|\mu\!\cdot\!\epsilon_i|\le\|\mu\|\|\epsilon_i\|\le\|\epsilon_i\|\le\delta$ and
$|\epsilon_i\cdot\epsilon_j|\ge -\|\epsilon_i\|\|\epsilon_j\|\ge -\delta^2$
Then
\[
\begin{split}
x_i\cdot x_j &\ge 1 - \|\epsilon_i\| - \|\epsilon_j\| + \epsilon_i\cdot\epsilon_j \\
&\ge 1 - 2\delta + \epsilon_i\cdot\epsilon_j \\
&\ge 1 - 2\delta - \delta^2 \\
\end{split}
\]
And
\[
\begin{split}
&1 + 2 \mu\cdot\epsilon_i + \|\epsilon_i\|^2 \le 1 + 2\delta + \delta^2, \\
&1 + 2 \mu\cdot\epsilon_j + \|\epsilon_j\|^2 \le 1 + 2\delta + \delta^2
\end{split}
\]
Hence
\[
\|x_i\|\,\|x_j\| \le 1 + 2\delta + \delta^2
\]
Combining these two inequalities yields the uniform lower bound
\[
\cos\theta^{(x)}_{ij}
= \frac{x_i\cdot x_j}{\|x_i\|\,\|x_j\|}
\ge \frac{1-2\delta-\delta^2}{1+2\delta+\delta^2}
\]

Therefore, if
\[
\cos\theta^{(y)}_{ij} \le \frac{1-2\delta-\delta^2}{1+2\delta+\delta^2}
\]
we have $\cos\theta^{(y)}_{ij}\le\cos\theta^{(x)}_{ij}$. By Corollary 1 (monotonicity of $E[I]$ in the angles), this implies
\[
E[I^{(y)}]\ge E[I^{(x)}]
\]

\hfill\(\square\)

\subsection{Algorithm Tables}

We include SeqMark text generation in Algorithm \ref{alg:generate} and text detection in Algorithm \ref{alg:detect}

\begin{algorithm*}[!]
\caption{SeqMark text generation algorithm}
\label{alg:generate}

\SetKwProg{Fn}{def}{:}{end}

\SetKwInOut{Input}{Input}
\SetKwInOut{Params}{Params}
\SetKwInOut{Output}{Output}

\Input{input prompt $p$, number of sentences to generate $T$}
\Params{language model $P(\cdot)$, sentence encoder $E(\cdot)$ with embedding dimension $h$, manifold sample size $n$, maximum rejections $N_{\max}$, LSH dimension $d$, valid region ratio $\gamma \in (0,1)$ secret key $k$}
\Output{watermarked sequence $s_{1:T}$}

\BlankLine

\SetKwFunction{GENERATE}{GENERATE}
\SetKwFunction{SIG}{SIG}
\SetKwFunction{LSH}{LSH}

\Fn{\GENERATE{$p$, $T$}}{
\
\SetKwBlock{Init}{Initialization:}{}
\Init{
  Randomly initialize $d$ vectors $n^{(1)}...n^{(d)} \in \mathbb{R}^h$, to create $2^d$ semantic subspaces.
}

\BlankLine
\For{$t\leftarrow 1$ \KwTo $T$}{

  \tcp{Compute valid and blocked regions}
    Compute the LSH signature of the previously generated sentence, $\mathrm{SIG}(s_{t-1})$, and use $\big[\mathrm{SIG}(s_{t-1})\big]_{10} \cdot p$ as the seed to randomly divide the signature space $\{0,1\}^{d}$ into a \emph{valid region set} $G_t$ of size $\gamma\,2^{d}$ and a \emph{blocked region set} $R_t$ of size $(1-\gamma)\,2^{d}$.
    
  \tcp{Estimate mean embedding of high-quality manifold}
  \For{$i\leftarrow 1$ \KwTo $n$}{
    $s_t^{(i)} \sim P(\cdot \mid p)$\;
    $c_i \leftarrow E(s_t^{(i)})$\;
  }
  $\bar c_t \leftarrow \dfrac{1}{n}\sum_{i=1}^n c_i$\;

  \tcp{Rejection sampling with mean-centering}
  \For{$r\leftarrow 1$ \KwTo $N_{\max}$}{
    $s_t^\ast \sim P(\cdot \mid p)$\;
    $u^\ast \leftarrow E(s_t^\ast) - \bar c_t$\;
    $b^\ast \leftarrow LSH(u^\ast)$\;
    \If{$b^\ast \in G_t$}{
      $s_t \leftarrow s_t^\ast$\;
      \textbf{break}\;
    }
  }
  \If{$r = N_{\max}$}{
    set $s_t$ to the last candidate observed\;
  }

  \tcp{Concatenate newly generated sentence into prompt}
  $p \leftarrow p || s_t$
}

\Return{$s_{1:T}$}
}

\BlankLine
\tcp{Subroutines}
\Fn{\SIG{$s$}}{
    $v \leftarrow E(s)$ \tcp*{embedding of sentence $s$}
    $c \leftarrow LSH(v)$ \tcp*{LSH signature of the embedding}
    \Return{$c$}\;
}

\Fn{\LSH{$v$}}{
    Initialize $c \leftarrow []$\;
    \For{$i \leftarrow 1$ \KwTo $d$}{
        $c \leftarrow c \,\|\, \left( n^{(i)} \cdot v > 0 \right)$\tcp*{Append digit}
    }
    \Return{$c$}
}
\end{algorithm*}

\begin{algorithm*}[!t]
\caption{SeqMark text detection algorithm}
\label{alg:detect}

\SetKwProg{Fn}{def}{:}{end}

\SetKwInOut{Input}{Input}
\SetKwInOut{Params}{Params}
\SetKwInOut{Output}{Output}

\Input{input prompt $p$, text $T = (s_1, \dots, s_M)$}
\Params{language model $P(\cdot)$, sentence encoder $E(\cdot)$ with embedding dimension $h$, 
LSH dimension $d$, valid region ratio $\gamma \in (0,1)$, 
secret key $k$}
\Output{number of detected watermarked sentences $D$}

\SetKwFunction{DETECT}{DETECT}

\BlankLine

\Fn{\DETECT{$T$}}{

\SetKwBlock{Init}{Initialization:}{}
\Init{
  Randomly initialize the same $d$ hyperplane normals 
  $n^{(1)}, \dots, n^{(d)} \in \mathbb{R}^h$ using key $k$\;
  $D$ $\leftarrow 0$\; \tcp{Number of detected sentences}
}

\BlankLine

\For{$t \leftarrow 1$ \KwTo $M$}{

  \tcp{Re-compute valid and blocked regions}
    Compute the LSH signature of the previously generated sentence, $\mathrm{SIG}(s_{t-1})$, and use $\big[\mathrm{SIG}(s_{t-1})\big]_{10} \cdot p$ as the seed to randomly divide the signature space $\{0,1\}^{d}$ into a \emph{valid region set} $G_t$ of size $\gamma\,2^{d}$ and a \emph{blocked region set} $R_t$ of size $(1-\gamma)\,2^{d}$.
    
  \tcp{Re-estimate mean embedding of high-quality manifold}
  \For{$i\leftarrow 1$ \KwTo $n$}{
    $s_t^{(i)} \sim P(\cdot \mid p)$\;
    $c_i \leftarrow E(s_t^{(i)})$\;
  }
  $\bar c_t \leftarrow \dfrac{1}{n}\sum_{i=1}^n c_i$\;

  \tcp{Check the current sentence}
  $v_t \leftarrow E(s_t) - \bar c_t$\;
  $b_t \leftarrow LSH(v_t)$\;

  \If{$b_t \in G_t$}{
      $D \leftarrow D + 1$\;
  }
  \tcp{Concatenate sentence into prompt for next round}
  $p \leftarrow p || s_t$
}

\Return{$D$}
}

\end{algorithm*}

\begin{table*}[ht]
\centering
\small
\setlength{\tabcolsep}{1.5pt}
\renewcommand{\arraystretch}{1.15}
\begin{tabular}{@{}l l l l l l l l@{}}
\toprule
\textbf{Dataset} & \textbf{Task}  & \textbf{Language} & \textbf{\#Data} & \textbf{LLM} & \textbf{Sentence Encoder} & \textbf{Text Quality Metric}\\
\midrule
WMT19 & Translation  & De-En & 3000 & ALMA-7B & LABSE & COMET\\
XSum & Summarization & En & 1000 & Llama-2-7B-Instruct & all-mpnet-base-v1 & ROUGE-L / COMET \\
MBPP & Code Generation & En & 500 & CodeLlama-7B-Instruct & embeddinggemma-300M & pass@k \\
C4 & Generation & En & 100 & Llama-2-7B-Instruct & all-mpnet-base-v1 & Perplexity \\
WMT23 & Long-form MT& De-En & 100  & Gemma-2-4B-Instruct & LABSE & BLEU / COMET \\
\bottomrule
\end{tabular}
\caption{Datasets, tasks, models, and evaluation metrics used in our experiments.}
\label{tab:datasets}
\end{table*}

\begin{table*}[!th]
\centering
\scalebox{0.80}{
\begin{tabular}{l}
\toprule
\textbf{Translation Prompt} \\
\midrule
Translate the following text from German to English: \\
German: München 1856: Vier Karten, die Ihren Blick auf die Stadt verändern \\
English: \\
\midrule
\textbf{Summarization Prompt} \\
\midrule
For the following article, write a one-sentence summary: ```The ex-Reading defender denied fraudulent trading charges \\ 
relating to the Sodje Sports Foundation - a charity to raise money for Nigerian sport.
Mr Sodje, 37, is jointly charged \\ with elder brothers Efe, 44, Bright, 50 and Stephen, 42. Appearing at the Old Bailey earlier, all four denied the offence.\\
The charge relates to offences which allegedly took place between 2008 and 2014. Sam, from Kent, Efe and Bright, of  \\ Greater Manchester, and Stephen, from Bexley, are due to stand trial in July. They were all released on bail. \\
Summary:\\
\midrule
\textbf{Code Generation Prompt} \\
\midrule
Write a python function to remove first and last occurrence of a given character from the string. \\
Directly generate a Python function only.
Your code should satisfy these tests: \\
\texttt{assert remove\_Occ(hello,l) == heo}\\ 
\texttt{assert remove\_Occ(abcda,a) == bcd} \\
\texttt{assert remove\_Occ(PHP,P) == H} \\
Solution: \\
\bottomrule
\end{tabular}
}
\caption{Examples of prompt templates used in this work.}
\label{tab:prompts}
\end{table*}

\begin{table*}[!h]
\centering
\scalebox{1.0}{
\begin{tabular}{l}
\toprule
\textbf{Source}: München 1856: Vier Karten, die Ihren Blick auf die Stadt verändern \\ 
\textbf{Target}: Munich 1856: Four maps that will change your view of the city \\ 
\textbf{NS-Watermark}: Munich 1856: Four \textcolor{red}{\underline{Eng}ravings} that Change Your View of the City \\ 
\textbf{\ours}: Munich 1856: Four maps that change your view of the city  \\
\midrule 
\textbf{Source}: Kleingärtner bewirtschaften den einstigen Grund von Bauern. \\
\textbf{Target}: Allotment holders cultivate the soil of former farmers. \\
\textbf{NS-Watermark}: Allotment gardeners cultivate the former \textcolor{red}{\underline{from}} of farmers. \\
\textbf{\ours}: Allotment holders cultivate the former fields of farmers. \\
 \midrule 
\textbf{Source}: Es nervt, wenn Landkarten nicht aktuell sind. \\ 
\textbf{Target}: It is annoying when geographical maps are not up-to-date. \\
\textbf{NS-Watermark}: It's annoying when maps \textcolor{red}{\underline{being}} out of date. \\
\textbf{\ours}: It annoys me when maps are not up-to-date. \\
\bottomrule
\end{tabular}
}
\caption{Three qualitative examples from WMT19.  We include the source sentence, target sentence, NS-Watermark and SeqMark translations. We use underline to denote the \underline{watermarked tokens}. Interestingly, the watermarked tokens in NS-Watermark are often the odd ones out in the translation (highlighted in \textcolor{red}{red}). We further note that NS-Watermark only uses beam search for decoding, which is fundamentally different from other approaches (including ours) that utilize token sampling.}
\label{tab:ns_watermark_qualitative}
\end{table*}


\end{document}